\documentclass[12pt]{JHEP3}
\usepackage{amsfonts}
\usepackage{amscd}
\usepackage{amssymb}
\usepackage{amsmath}
\usepackage{epsfig}
\usepackage{latexsym}

\def \be  {\begin{equation}}
\def \ee  {\end{equation}}
\def \ba  {\begin{eqnarray}}
\def \ea  {\end{eqnarray}}
\def \bb  {\begin{thebibliography}}
\def \eb  {\end{thebibliography}}
\def \lab #1 {\label{#1}}
\def \ni   {\noindent}

\newcommand\cA{\mathcal{A}}

\newcommand{\rf}{*}

\newcommand\lb{\lambda}

\newcommand\Nsq{\mathrm{N}^2\mathrm{MHV}}

\newcommand\MHV{\mathrm{MHV}}

\newcommand\cN{\mathcal{N}}

\newcommand\C {\mathbb{C }}

\newcommand\CP {\mathbb{CP}}

\newcommand\rd{\mathrm{d}}

\newcommand\im{\mathrm{i}}
\newcommand\la{\langle}
\newcommand\ra{\rangle}

\newcommand\delbar{\bar{\partial}}

\newcommand{\bea}{\begin{eqnarray}\label}
\newcommand{\eea}{\end{eqnarray}}

\title{MHV Diagrams in Momentum Twistor Space}

\author{Mathew Bullimore\\
	Rudolf Peierls Centre for Theoretical Physics,\\
	1 Keble Road, Oxford, OX1 3NP,
	United~Kingdom}
	
\author{Lionel Mason\\
	The Mathematical Institute,\\
	24-29 St.~Giles', Oxford, OX1 3LB,
	United~Kingdom}

\author{David Skinner\\
	Perimeter Institute for Theoretical Physics,\\ 
	31~Caroline~St., Waterloo, ON, N2L 2Y5, 
	Canada}

\abstract{ We show that there are remarkable simplifications
        when the MHV diagram formalism for $\cN=4$ super Yang-Mills is
        reformulated in momentum twistor space.  The vertices are
        replaced by unity while each propagator becomes a dual
        superconformal `R-invariant' whose arguments may be read off
        from the diagram, and include an arbitrarily chosen
        reference twistor. The momentum twistor MHV rules generate a
        formula for the full, all-loop planar integrand for the super
        Yang-Mills S-matrix that is manifestly dual superconformally invariant up
        to the choice of a reference twistor. We give a general proof
        of this reformulation and illustrate its use by computing the
        momentum twistor NMHV and N$^2$MHV tree amplitudes and the
        integrands of the MHV and NMHV 1-loop and the MHV 2-loop
        planar amplitudes.

}
\begin{document}

%%%%%%%%%%%%%%%%%%%%%%%%%%%%%%%%%%%%%%%%%%%%%%%%%%%
%%%%%%%%%%%%%%%%%%%%%%%%%%%%%%%%%%%%%%%%%%%%%%%%%%%

\section{Introduction}
\label{sec:intro}

The MHV diagram formalism~\cite{Cachazo:2004kj} is a set of momentum space Feynman-like rules for calculating scattering amplitudes in supersymmetric gauge theories. In these rules, the propagator is the scalar propagator $1/p^2$, while the vertices are essentially MHV amplitudes, extended off-shell by means of a `reference spinor'. The MHV formalism provides a substantial simplification over the use of conventional Feynman diagrams and has lead to direct calculations of a number of tree and loop amplitudes~\cite{Bena:2004ry, Brandhuber:2004yw, Brandhuber:2005kd}.  It
is now well established at tree level~\cite{Risager:2005vk,Elvang:2008na,Elvang:2008vz} and for supersymmetric
gauge theories at 1-loop~\cite{Brandhuber:2005kd}. It has also been extended to many other gauge theories with different forms of matter, see {\it e.g.}~\cite{Boels:2007qn, Boels:2007pi,Boels:2008ef,Glover:2008tu,Buchta:2010qr}.  MHV diagrams are the axial gauge Feynman diagrams of a supersymmetric Yang Mills action in twistor space, which in a different gauge is equivalent to the standard space-time action~\cite{Boels:2007qn} and has also been obtained from the light cone gauge space-time action in various ways~\cite{Mansfield:2005yd,Ettle:2006bw,Lovelace:2010ev}.

In a separate line, dual superconformal invariance has proved a powerful tool in the study of scattering amplitudes in planar $\cN=4$ SYM~\cite{Drummond:2008vq,Brandhuber:2008pf}, leading to a complete formula for the amplitude at tree-level~\cite{Drummond:2008cr} and many new insights at loop level~\cite{Drummond:2008bq,Mason:2009qx,Mason:2009sa,Kaplan:2009mh}.  It can be motivated as arising from T-duality in the AdS/CFT approach to calculating amplitudes at strong coupling~\cite{Alday:2007hr,Alday:2009dv,Alday:2010vh} as areas of minimal surfaces in AdS. This string based approach has a field theory limit in which amplitudes are obtained as the correlation function of a Wilson loop around a null polygon in region momentum space~\cite{Drummond:2007aua,Drummond:2007au,Brandhuber:2007yx,Drummond:2008aq,Anastasiou:2009kna}.

The spinors of the dual conformal group have been introduced to this context by Hodges~\cite{Hodges:2009hk} and termed `momentum twistors' because the space-time that they correspond to is region momentum space. The null polygon in region momentum space is dual to a polygon in momentum twistor space $\CP^{3|4}$ and is determined by $n$-twistors.  The correspondence between the $n$ null momenta and $n$ momentum twistors is essentially a purely algebraic change of variables but serves to manifest dual superconformal invariance.  Hodges~\cite{Hodges:2009hk} gave a geometric interpretation of NMHV tree diagrams and~\cite{Mason:2009qx} gave a Grassmannian formulation of tree amplitudes and leading singularities that manifests dual superconformal invariance and is equivalent to the Grassmannian formulation of~\cite{ArkaniHamed:2009dn} in usual twistor space.

The main purpose of this article is to reformulate the MHV diagram rules into momentum twistor space.  We find that the reformulation does indeed simplify the formulae and manifest superconformal invariance as much as could be expected.  The original MHV diagram formalism violates even Lorentz invariance through the choice of a reference spinor.  In our momentum twistor formulation, this becomes a choice of reference twistor.  But for that choice, the formalism becomes dual superconformal invariant.  The Feynman-like rules for constructing an integrand from an MHV diagram become simply the association of a dual superconformal `R-invariant' whose arguments are the reference twistor and certain other twistors that can be read off the diagram near the propagator.  The vertices are replaced by unity.   Thus the formalism gives a direct algorithm to calculate the all-loop integrand for $N=4$ super Yang-Mills amplitudes.  This algorithm is particularly efficient for amplitudes of low MHV degree and we give a number of examples at both tree and loop-level.

Of course these loop integrands must be integrated in order to obtain the amplitudes. These integrals are divergent and need to be regularised.  The traditional choice is to use dimensional regularisation, but this does violence to the twistor correspondence.  However, the Coulomb branch regularisation proposed in~\cite{Alday:2009zm} -- a natural, physically motivated way to regulate the infra-red divergences of amplitudes -- meshes very naturally with momentum twistor geometry, as found in~\cite{Hodges:2010kq,Mason:2010pg,Drummond:2010mb, Alday:2010jz}.

\medskip

There are a number of avenues for further investigation that follow on from these ideas that we plan to address in subsequent papers.   In particular, in a companion paper~\cite{Wilson-Loop} we make contact with the Wilson loop story in dual conformal space-time by reformulating the Wilson loop in momentum twistor space and calculating its correlation function using the twistor action in an axial gauge.  This leads precisely to the MHV formalism for scattering amplitudes in momentum twistor space obtained in this paper, but with the corresponding Feynman diagrams being the planar duals of the MHV diagrams for the scattering amplitudes.

The MHV diagram formalism is widely believed to give the correct answers for reasonably general supersymmetric gauge theories to all loop orders.  However, this has only been proved systematically at tree-level~\cite{Risager:2005vk,Elvang:2008vz} and 1-loop~\cite{Brandhuber:2005kd} (unless one is prepared to take the twistor action argument to be a proof).   The work of this paper together with the recent work on BCFW recursion relations for loop integrands~\cite{Arkani-Hamed:2010kv, Boels:2010nw} suggests that recursion in momentum twistor  space might lead to a proof for the all-loop integrand.  Indeed, Risager recursion in momentum twistor space is based on a shift of the external momentum twistors along the reference twistor and should be straightforward to implement in this context.  

This version of the MHV formalism leads to formul\ae\ that are very close to those obtained by BCFW recursion.  The fact that R-invariants are naturally associated to the propagators in MHV diagrams is suggestive also of a similar diagram formalism for the formul\ae\ for amplitudes obtained by BCFW recursion in terms of R-invariants.  Indeed this is clear for the Drummond and Henn formula for the tree level $\cN=4$ super Yang-Mills amplitudes, and the pattern seems likely to continue for the formul\ae\ obtained by BCFW recursion for the loop integrand~\cite{Arkani-Hamed:2010kv,Boels:2010nw}.

\medskip

The paper is structured as follows.    We give brief introductions to momentum twistors in section~\ref{sec:momtwistors} and to the MHV diagram formalism in section~\ref{sec:CSW}.  In  section~\ref{sec:trees}, we describe and prove the momentum twistor space MHV rules for an arbitrary $\cN=4$ tree amplitude, and  illustrate their use in the NMHV, N$^2$MHV and N$^3$MHV cases.  The main difficulty here is to keep track of how external twistors need to be shifted when placed into R-invariants associated to sequences of adjacent propagators.  In section~\ref{sec:MHVloops} we extend the formalism to  1-loop amplitudes, again illustrating their use with MHV and NMHV examples.   At 1-loop, not only are propagators adjacent,  but one also needs to incorporate a pair of twistors that describe the loop momentum.    We give a comparison to unitarity methods in section~\ref{sec:boxes} to verify the independence from the reference twistor in the MHV case (this is of course known independently from the work of \cite{Brandhuber:2005kd}).   Finally, we explain the extension to all loops in section~\ref{sec:allloops}.

%%%%%%%%%%%%%%%%%%%%%%%%%%%%%%%%%%%%%%%%%%%%%%%%%%%
%%%%%%%%%%%%%%%%%%%%%%%%%%%%%%%%%%%%%%%%%%%%%%%%%%%

\section{Momentum Twistors for $\cN=4$ super Yang-Mills}
\label{sec:momtwistors}

An on-shell $\cN=4$ supermultiplet
\be
\label{supermultiplet}
	\Phi(\lambda,\tilde\lambda,\eta) = G^+(\lambda,\tilde\lambda) + \eta^a\Gamma_a(\lambda,\tilde\lambda) + \cdots
	+\frac{\epsilon_{abcd}}{4!}\eta^a\eta^b\eta^c\eta^dG^-(\lambda,\tilde\lambda)\ .
\ee
depends on bosonic spinor momenta\footnote{
		$A=0,1$ and $A'=0',1'$ are anti self-dual and self-dual Weyl spinor indices, while
		$a=1,\dots,4$ is an $R$-symmetry index. We typically suppress these
		indices in what follows.} 
$(\lambda_A, \tilde\lambda_{A'})$ and a fermionic variable $\eta^a$ that the counts the helicity of the component fields. In the planar sector of $n$-particle scattering amplitudes, the colour-ordering allows us to naturally encode the $n$ such supermomenta into $n$ region supermomenta $(x^{AA'}_i,\theta^{Aa}_i)$ defined for $i=1,\ldots,n$ up to translation by
\be
\label{regionmom}
	x_i - x_{i+1}=\lambda_i\tilde\lambda_i \qquad \theta_i- \theta_{i+1}=\lambda_i\eta_i\ ,
\ee 
with $(x_{n+1},\theta_{n+1})\equiv (x_1,\theta_1)$. It is clear that, in terms of these variables, the amplitude is invariant under a common translation of the $(x_i,\theta_i)$ as well as Lorentz transformations.  Working with region momenta also ensures that supermomentum conservation
\be 
	\sum_i \lambda_i\tilde\lambda_i =0\qquad \sum_i \lambda_i\eta_i =0
\ee 
is automatic. 

If one pulls out an overall factor of the MHV tree amplitude, writing
\be
	A(1,\ldots,n) = A_{\rm MHV}^{(0)}(1,\ldots,n)\,M(1,\ldots,n)
\label{Mdef}
\ee
for an arbitrary amplitude, one finds that tree-level ratio functions $M$ are invariant under a \emph{dual} superconformal group~\cite{Drummond:2008vq}. This group acts on the region momenta $(x_i,\theta_i)$ in exactly the same manner as the usual superconformal group acts on space-time. At loop level the ratio functions may be represented as integrals of rational functions. While the loop integrals themselves are divergent and require regularisation, the \emph{integrands} of planar amplitudes are again found to be invariant under the dual superconformal group~\cite{Arkani-Hamed:2010kv}. The closure of the ordinary and dual superconformal algebras was shown in~\cite{Drummond:2009fd} to be the full Yangian Y$[\frak{psl}(4|4;\C)]$ associated to the superconformal algebra, and is believed to be a symmetry of planar $\cN=4$ SYM.

Twistors are the spinors of the (super)conformal group, so it is natural to construct a twistor space for the region momenta. This was introduced by Hodges~\cite{Hodges:2009hk}, who named it \emph{momentum twistor space} to distinguish it from the standard twistor space associated to the ordinary superconformal group. With $\cN=4$ supersymmetry, momentum twistor space is a copy of $\C^{4|4}$ and may be described by four bosonic and four fermionic coordinates $Z^I = (Z^\alpha;\chi^a) = (\lambda_A,\mu^{A'};\chi^a)$. The key property of this space is that the (complexification of the four-fold cover of the) dual superconformal group SL$(4|4;\C)$ acts linearly, so that a twistor $Z$ is in the fundamental representation of the dual superconformal group.

We now give a brief introduction to (momentum) twistor geometry. More complete introductions may be found in~\cite{Hodges:2009hk,Mason:2009qx}, or indeed the standard textbooks~\cite{HuggettTod,Penrose:1986ca}. The region supermomenta determine $n$ momentum twistors $Z_i$ up to scale via the incidence relation
\be
\label{incidence}
	Z_i = (\lambda_{iA}\,,\,  x_i^{AA'}\lambda_{iA}\, ;\,  \theta^{Aa}_i\lambda_{iA})\,.
\ee
Using \eqref{regionmom}  we also have 
\be
\label{incidence+1}
	Z_i=(\lambda_{iA}\,,\, x_{i+1}\lambda_{iA}\,;\, \theta^{Aa}_{i+1}\lambda_{iA})
\ee 
and these equations together determine the region supermomenta momenta:
\be
\label{reconstruction}
	x_i=\frac{\mu_{i-1}\lambda_i - \mu_i\lambda_{i-1}}{\la i,i\!-\!1\ra}\,\qquad
	\theta_i = \frac{\chi_{i-1}\lambda_i - \chi_i\lambda_{i-1}}{\la i,i\!-\!1\ra}\,,
\ee
where the angle bracket with two arguments is the usual SL$(2;\C)$-invariant spinor product $\la i,j\ra \equiv \epsilon^{AB}\lb_A^{(i)}\lb_B^{(j)}$.  When needed, the supermomenta themselves can be reconstructed from~\eqref{regionmom} as
\be
\label{recon-mom}
\begin{aligned}
	\tilde\lambda_i &=\frac{ \mu_{i-1} \la i\, i\!+\!1\ra + \mu_i\la i\!+\!1\, i\!-\!1\ra +\mu_{i+1} \la i\!-\!1\, i\ra}
				{\la i\!-\!1, i\ra\,\la i, i\!+\!1\ra}\\
	\eta_ i& = \frac{\chi_{i-1} \la i\, i\!+\!1\ra + \chi_i\la i\!+\!1\, i\!-\!1\ra +\chi_{i+1} \la i\!-\!1\, i\ra}
				{\la i\!-\!1, i\ra\,\la i, i\!+\!1\ra}\, .
\end{aligned}
\ee

\begin{figure}[t]
	\centering
	\includegraphics[height=40mm]{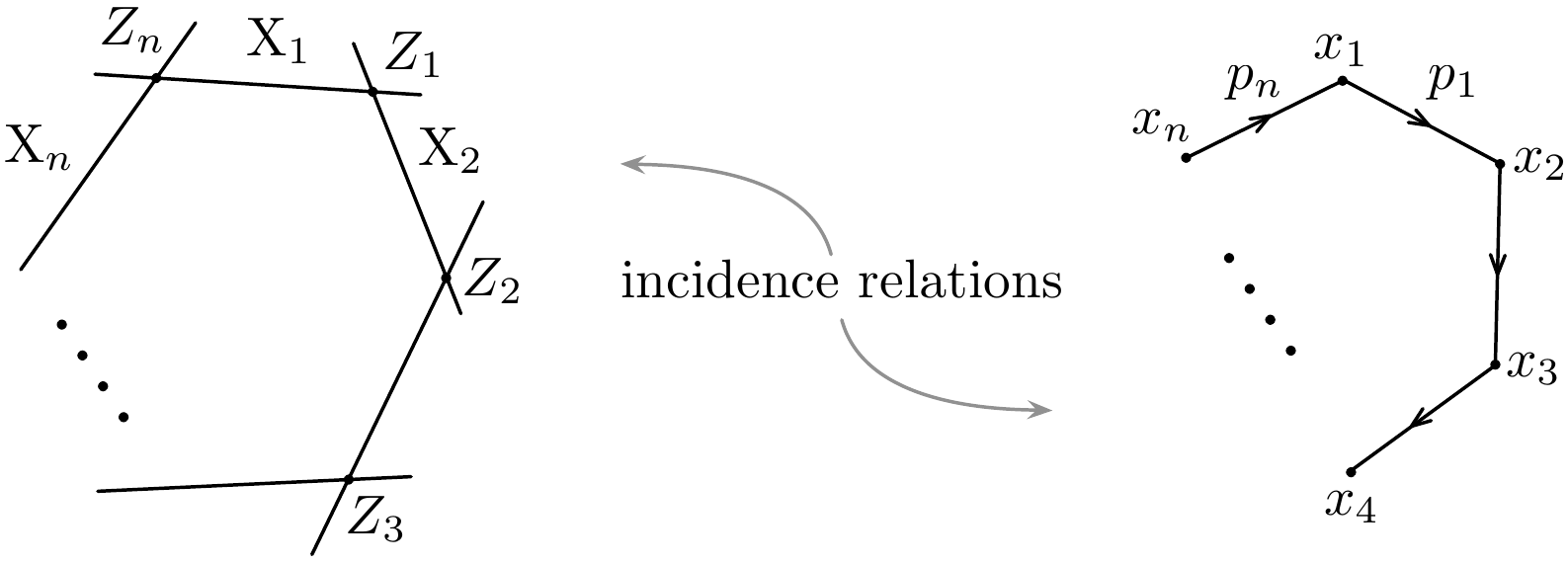}
	\caption{The momentum twistor correspondence.}
	\label{fig:momtwistor}
\end{figure}

These equations do not fix the scale of the momentum twistors, so they determine only the sequence of $n$ points $[Z_i]$ in the \emph{projective} space $\CP^{3|4}=(\C^{4|4}-\{0\})/\C^*$.  Triviality under overall rescaling reduces the action of SL$(4|4;\C)$ to PSL$(4|4;\C)$.  Geometrically, equations~\eqref{incidence}-\eqref{reconstruction} simply state that the region momentum $x_i$ corresponds to a complex line X$_i\cong\CP^1$ in projective twistor space, and that this line is the \emph{join} $(i\!-\!1,i)$ of the twistors $Z_i$ and $Z_{i-1}$.  In consequence, the null polygon in region momentum space determines and is determined by a polygon in momentum twistor space, as shown in figure~\ref{fig:momtwistor}.

\medskip

Momentum twistors have been usefully employed to study the Grassmannian residue formula~\cite{Mason:2009qx} in the context of leading singularities, to formulate a BCFW-style recursion relation for loop integrands~\cite{Arkani-Hamed:2010kv} and to evaluate certain one- and two-loop integrals themselves~\cite{Hodges:2010kq,Mason:2010pg,Drummond:2010mb}. The advantage of the framework is not just the manifest dual superconformal symmetry, but also the fact that the momentum twistors $Z_i$ may be freely prescribed, in contrast to the region momenta that are constrained by $(x_i-x_{i+1})^2=0$, or the original momenta $p_i$ that are additionally constrained to sum to zero.

One of the first places these benefits are felt -- and one that will be important throughout this paper -- comes from rewriting the dual superconformal invariant~\cite{Drummond:2008vq}
\be
	R_{n;ij} = \frac{\la i\!-\!1,i\ra\,\la j\!-\!1,j\ra\ \delta^{0|4}(\la n|x_{nj}x_{ji}|\theta_{in}\ra + \la n|x_{ni}x_{ij}|\theta_{jn}\ra)}
			{x_{ij}^2 \la n|x_{nj}x_{ji}|i\ra\,\la n|x_{nj}x_{ji}|i\!-\!1\ra\,\la n|x_{ni}x_{ij}|j\ra\,\la n|x_{ni}x_{ij}|j\!-\!1\ra}
\label{momR}
\ee
in terms of momentum twistors. This was first done in~\cite{Mason:2009qx} with the result
\be
	R_{n;ij} = \frac{\delta^{0|4}\!\left(\la n,i\!-\!1,i,j\!-\!1\ra \chi_j + \hbox{cyclic}\right)}
			{\la i\!-\!1,i,j\!-\!1,j\ra\,\la i,j\!-\!1,j,n\ra\,\la j\!-\!1,j,n,i\!-\!1\ra\,\la j,n,i\!-\!1,i\ra\,\la n,i\!-\!1,i,j\!-\!1\ra}\,,
\label{Rdef}
\ee
where we have introduced the SL$(4;\C)$-invariant skew product of four (bosonic) twistors 
\be
	\la a,b,c,d\ra \equiv \epsilon_{\alpha\beta\gamma\delta}Z_a^\alpha Z_b^\beta Z_c^\gamma Z_d^\delta\,.
\ee
Each of the five factors in the denominator of~\eqref{momR} directly corresponds to factor in the denominator of~\eqref{Rdef}, for example,
\be
	x_{ij}^2 = \frac{\la i\!-\!1,i,j\!-\!1,j\ra}{\la i\!-\!1,i\ra\,\la j\!-\!1,j\ra} \qquad
	\la n|x_{nj}x_{ji}|i\ra = \frac{\la i,j\!-\!1,j,n\ra}{\la j\!-\!1,j\ra}
\ee
as follows directly from the incidence relations~\eqref{incidence}. Thus, $R_{n;ij}$ has a simple pole when any four of the five twistors on which it depends become coplanar.

The form~\eqref{Rdef} for the R invariant is manifestly dual conformal. As shown in~\cite{Mason:2009qx}, we can go further and write 
\be
	R_{n;ij} = \int_{\CP^4}\frac{{\rm D}^4C}{C_1C_2C_3C_4C_5} \,
	\bar\delta^{4|4}\!\left(C_1Z_n + C_2Z_{i-1}  + C_3Z_i + C_4 Z_{j-1} + C_5Z_j\right)
\label{Rintegral}
\ee	
in terms of the five supertwistors. Equation~\eqref{Rintegral} makes dual \emph{super}conformal invariance manifest, and is the basic building block of the momentum twistor Grassmannian. Either form also makes clear that a generic R invariant is really a totally antisymmetric function $[\ ,\ ,\ ,\ ,\ ]$ that may be defined for five arbitrary momentum twistors. Using this notation, we have $R_{n;ij} = [n,i\!-\!1,i,j\!-\!1,j]$, while the multi-index R invariants $R_{n;ab;cd;\ldots;ef}$ introduced in~\cite{Drummond:2008cr} are again simply examples of $[\ ,\ ,\ ,\ ,\ ]$ evaluated on certain shifted twistors, as explained in~\cite{Mason:2009qx}. We shall meet these R invariants throughout the paper.

%%%%%%%%%%%%%%%%%%%%%%%%%%%%%%%%%%%%%%%%%%%%%%%%%%%
%%%%%%%%%%%%%%%%%%%%%%%%%%%%%%%%%%%%%%%%%%%%%%%%%%

\section{The MHV formalism}
\label{sec:CSW}

MHV diagrams~\cite{Cachazo:2004kj} are the Feynman diagrams of the twistor action~\cite{Mason:2005zm,Boels:2006ir,Boels:2007qn} in an axial gauge.  They consist of $n$-valent, local space-time vertices connected by propagators.  These vertices are based on the MHV superamplitudes, given in momentum space by the famous Parke Taylor formula
\be
	A_{\rm MHV}^{(0)} =  \frac{\delta^{4|8}\!\left(\sum \lambda_i\tilde\lambda_i\right)}
	{\la 12\ra\la23\ra\,\cdots\,\la n1\ra}\, .
\label{MHVtreeamp}
\ee
To promote~\eqref{MHVtreeamp} to a vertex, one must be able to assign an unprimed spinor $|\lambda\ra$ to each leg, even off-shell. The prescription of Cachazo, Svr{\v c}ek and Witten is to pick an arbitrary, fixed reference spinor $\iota^{A'}$ and define
\be
	\lambda_A \equiv p_{AA'} \iota^{A'}
\label{CSWprescription}
\ee
for every off-shell momentum $p$ in the diagram.  The $|\lb\ra$ spinors are then used in~\eqref{MHVtreeamp} for off-shell legs in exactly the same manner as the spinor $|i\ra$ is used for the $i^{\rm th}$ on-shell leg.  In the $\cN=4$ theory, one also assigns a Grassmann variable $\eta$ to each leg, on-shell or otherwise. These fermionic variables are used in the Grassmann $\delta^{0|8}$-function in~\eqref{MHVtreeamp}. After multiplying~\eqref{MHVtreeamp} by a factor of $1/n$ to account for possible rotations of the vertex before connecting it into a diagram, this prescription promotes~\eqref{MHVtreeamp} to an off-shell `MHV vertex'.

From the perspective of the twistor action, the choice of CSW reference spinor is a choice of axial gauge direction, while the whole infinite series of vertices are naturally packaged as
\be
	\int\rd^{4|8}x\,\ln\det(\delbar+\cA|_{\rm X}) 
	= \int \rd^{4|8}x\  {\rm Tr}\left(\ln \delbar^{-1}|_{\rm X} +
	\sum_{n=1}^\infty \,\frac{1}{n} (\delbar^{-1}\!\cA_1\,\delbar^{-1}\!\cA_2\,\cdots\delbar^{-1}\!\cA_n)\right)
\ee
where X is the line is twistor space (not momentum twistor space!) corresponding to the space-time point $x$ (not region momentum!). 

MHV diagrams are formed by joining MHV vertices together with propagators. Each propagator that carries off-shell momentum $p$ contributes a factor of $1/p^2$ to the diagram, as again follows from the twistor space propagator in axial gauge\footnote{This will be discussed further in the forthcoming papers~\cite{Wilson-Loop,withTim}.}~\cite{Boels:2007qn}. To account for the possible members of the supermultiplet that are travelling along the propagator, 
one integrates over the fermionic variables $\eta$ associated to each propagator. For example, the simplest non-trivial diagram connects two MHV vertices with a single propagator and generates the expression
\be
	\int \rd^4\eta\  A_{\MHV}^{(0)}(\ldots,\{\lb,\eta\})\frac{1}{p^2}A_{\MHV}^{(0)}(\{\lb,\eta\},\ldots)\, . 
\end{equation}
in momentum space.

%%%%%%%%%%%%%%%%%%%%%%%%%%%%%%%%%%%%%%%%%%%%%%%%%%%
%%%%%%%%%%%%%%%%%%%%%%%%%%%%%%%%%%%%%%%%%%%%%%%%%%

\section{MHV tree diagrams in momentum twistor space} 
\label{sec:trees}

In this section we show that the momentum twistor expression for a tree-level MHV diagram can be obtained by replacing the vertices by unity and assigning an R-invariant to each propagator in the diagram. An MHV diagram that contributes to a tree-level N$^k$MHV amplitude has $k$ propagators, and so yields a product of $k$ R-invariants.

We first state the momentum twistor MHV rules in sufficient generality to handle an arbitrary tree-level MHV diagram. We then illustrate the use of this rule in the simple examples of the NMHV and N$^2$MHV tree amplitudes, and a particular diagram contributing the N$^3$MHV tree, before proceeding to give the general proof in section~\ref{sec:gentrees}.

%%%%%%%%%%%%%%%%%%%%%%%%%%%%%%%%%%%%%%%%%%%%%%%%%%%
%%%%%%%%%%%%%%%%%%%%%%%%%%%%%%%%%%%%%%%%%%%%%%%%%%%

\subsection{Momentum twistor MHV rules for tree amplitudes}
\label{sec:TreeRule}

\begin{figure}[t]
	\centering
	\includegraphics[height=37mm]{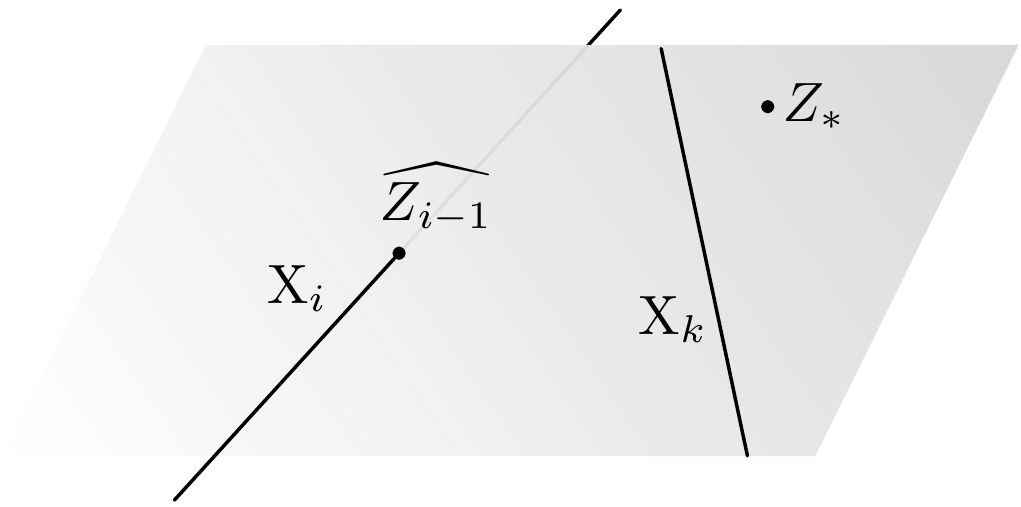}
	\caption{The shifted twistor $\widehat{Z_{i-1}}$ is the intersection of the line $(i\!-\!1,i)$ with the plane 
	$(\,\rf\,,k\!-\!1,k)$.}
	\label{fig:transversal}
\end{figure}

As will be proved below, the momentum twistor expression associated to a tree-level MHV diagram may be determined by the rule:
\begin{itemize}
	\item To each propagator separating region $x_i$ from region $x_j$, assign a factor of 
		$$[\,\rf\,,\widehat{i\!-\!1},i,\widehat{j\!-\!1},j]\,,$$ where $Z_\rf$ is an arbitrary reference momentum (super)twistor.
\end{itemize}
The momentum twistors $\widehat{i\!-\!1}$ and $\widehat{j\!-\!1}$ are determined separately for each propagator as follows:
\begin{itemize}
	\item If the external leg $i-1$ is attached to a vertex on which the given propagator ends, then 
		$\widehat{Z_{i-1}}=Z_{i-1}$ is unshifted.
		
	\item If the external leg $i\!-\!1$ is not directly attached to a vertex on which the given propagator ends 
		(so that one must traverse at least one further propagator to reach $i\!-\!1$), then $\widehat{Z_{i-1}}$ is the
		intersection of the line $(i\!-\!1,i)$ with the plane spanned by the reference twistor $Z_\rf$ and the line 
		$(k\!-\!1,k)$ associated to the propagator that is adjacent to the given one in the direction of $i\!-\!1$.
		
\end{itemize}
(The shift for $\widehat{j\!-\!1}$ is determined similarly.)  The shifted value $\widehat{i\!-\!1}$ is depicted geometrically in figure~\ref{fig:transversal} and may be written in components as
\be
\begin{aligned}
		\widehat{Z_{i-1}}&\equiv (i\!-\!1,i)\cap(\,\rf\,,k\!-\!1,k)\\
		&= \la\,\rf\,,k\!-\!1,k,i\!-\!1\ra Z_i - \la\,\rf\,,k\!-\!1,k,i\ra Z_{i-1}\,,
\end{aligned}
\ee 
fully supersymmetrically. The region $x_i$ always lies between the external states $i\!-\!1$ and $i$. Furthermore, there is always a unique path between any two external legs of a tree diagram, so the prescription to move `in the direction of $i\!-\!1$' is unambiguous.
Note also that there is no need to shift $k\!-\!1$ in determining the location of $\widehat{i\!-\!1}$ -- any such shift would move $k\!-\!1$ along the line $(k\!-\!1,k)$ and so does not affect the plane $(\,\rf\,,k\!-\!1,k)$.

In the above rule, an external leg is only ever shifted if it
\emph{precedes} the given propagator in the cyclic ordering, so that
$i\!-\!1$ and $j\!-\!1$ may be shifted,  whereas $i$ and $j$ never
are. This rule thus requires that we pick an orientation of each MHV
diagram, which below will be clockwise. There is an equivalent rule
for the opposite choice of orientation, where one instead assigns the
R invariant $[\,\rf\,,i\!-\!1,\hat{\imath},j\!-\!1,\hat{\jmath}]$ to a
propagator adjacent to regions $x_i$ and $x_j$, with $\hat{i} =
(i\!-\!1,i)\cap(\rf,l\!-\!1,\hat{l})$ in case $i$ is not attached to a
vertex on which our given propagator ends, and where $l\!-\!1$ and $l$
are the other external legs associated to the adjacent propagator in
the direction of $i$. 
%There is also a symmetric, though more redundant, choice\footnote{
%		Picking the point that is the intersection of the
%		transversal to ${\rm X}_i$, ${\rm X}_k$ and $Z_\rf$
%		with the 
%		line ${\rm X}_i$ clearly treats $i$ and $k$
%		asymmetrically, reflecting the overall choice of
%		orientation inherent in 
%		our rules. Although we will not exploit this here,
%		there is in fact a point on the transversal that
%		treats $i$ and $k$ 
%		symmetrically. This is the point $Z_{[ik]}$ that is
%		equianharmonic to $Z_\rf$ and the intersections with  
%		X$_i$ and X$_k$ ({\it i.e.}, the cross ratio of the
%		four points is $-1$).  In components,  
%		$Z_{[ik]}= 2\widehat{Z_{i\!-\!1}} + \frac{1}{2}({\rm
%		X}_i\cdot{\rm X}_k)\,Z_\rf$. 
%		}.
Indeed the correct result will be obtained even if we choose different
orientations for different sequences of adjacent propagators.  For the most part, we will content ourselves with the rule stated above.

\medskip

\ni Before proving these rules in general, we illustrate their use with some simple examples.

%%%%%%%%%%%%%%%%%%%%%%%%%%%%%%%%%%%%%%%%%%%%%%%%%%%
%%%%%%%%%%%%%%%%%%%%%%%%%%%%%%%%%%%%%%%%%%%%%%%%%%%

\subsection{NMHV tree amplitudes}
\label{sec:NMHVtree}

\begin{figure}[t]
	\centering
	\includegraphics[height=23mm]{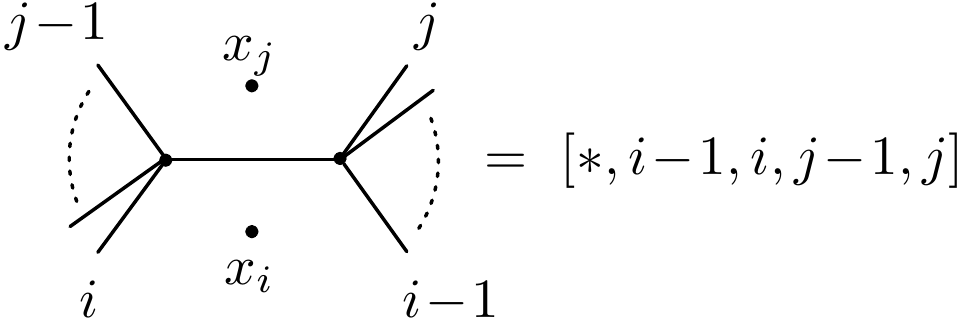}
	\caption{An MHV diagram contributing $[\rf,i\!-\!1,i,j\!-\!1,j]$ to the tree-level NMHV amplitude.}
	\label{fig:NMHVtree}
\end{figure}

Tree-level NMHV superamplitudes may be obtained from a sum of MHV diagrams each containing two MHV vertices and a single propagator, shown in figure \ref{fig:NMHVtree}. Since there is only one propagator, the lines adjacent to the propagator at each MHV vertex are inevitably external, so there are no shifts. Summing over all possible diagrams, the above rules give simply
\be
	M_{\rm NMHV} = \sum_{i<j}\,[\,\rf\,,i\!-\!1,i,j\!-\!1,j]
\label{NMHVtree1}
\ee
for the ratio of the NMHV tree amplitude to the MHV tree.

Let us check that this is the correct answer. Stripping away the overall super-momentum conserving $\delta$-function, each MHV diagram contributes
\be
	\int {\rm d}^4\eta\, A_{\MHV}(i,\ldots,j\!-\!1,\{\lb,\eta\})\frac{1}{(x_i-x_j)^2} A_{\MHV}(\{\lb,\eta\}, j,\ldots ,i\!-\!1)
\ee
on momentum space, where $\lb^A \equiv x_{ij}^{AA'}\iota_{A'}$ is the CSW prescription for the spinor associated to the off-shell momentum in the propagator, and $\iota_{A'}$ is an arbitrary reference spinor. Pulling out a factor of the $n$-particle Parke-Taylor denominator, we are left with 
\be
	\frac{\la i\!-\!1\, i\ra\la j\!-\!1\,j\ra}{x_{ij}^2\la\lb\,i\ra\la j\!-\!1\,\lb\ra\la\lb\,j\ra\la i\!-\!1\,\lb\ra}
	\times \int {\rm d}^4 \eta\ \delta^{0|8}(\theta_{ij} - \lb \eta)
\ee
where again $\la\lb| = [\iota|x_{ij}$. The fermionic integration is trivial to perform after factorising the fermionic $\delta$-function as
\be
	\delta^{0|8}(\theta_{ij} -\lb\eta)  
	=  \delta^{0|4} (\la\lb\,\theta_{ij}\ra ) \ \delta^{0|4}\!\left( \eta  -  \frac{\la \theta_{ij}\,1\ra}{\la\lb1\ra} \right)\,. 
\ee
We thus see that this particular MHV diagram contributes
\be 
	\frac{\la i\!-\!1\, i\ra\la j\!-\!1\,j\ra\, \delta^{0|4} ([\iota|x_{ij}|\theta_{ji}\ra) }
	{x_{ij}^2\,[\iota|x_{ij}|i\!-\!1\ra\,[\iota|x_{ij}|i\ra\,[\iota|x_{ij}|j\!-\!1\ra\,[\iota|x_{ij}|j\ra}
\label{NMHVdualco}
\ee
(times an overall $n$-particle MHV tree) to the NMHV tree. 

Introduce the auxiliary momentum twistor $Z_\rf=(0,\iota^{A'},0)$, the factors in the denominator of~\eqref{NMHVdualco} may be written in terms of momentum twistors as ({\it e.g.})
\be
	[\iota|x_{ji}|i\ra = \frac{\la \rf,j\!-\!1,j,i\ra}{\la j\!-\!1,j \ra} 
	\qquad\hbox{and}\qquad 
	x_{ij}^2 = \frac{\la i\!-\!1,i,j\!-\!1,j\ra}{\la i\!-\!1,i\ra\la j\!-\!1,j\ra}\,,
\label{transbos}
\ee
while the argument of the remaining fermionic $\delta^{0|4}$-function becomes
\be
\begin{aligned}
	{[i |x_{ij}|\theta_{ji}\ra} &= \la\rf,i\!-\!1,i,\left[j\!-\!1\ra,\chi_j\right] + \la\rf,j\!-\!1,j,\left[i\!-\!1\ra,\chi_i\right]\\
	&=\la\rf,i\!-\!1,i,j\!-\!1\ra\chi_j + \hbox{cyclic}\,,
\end{aligned}
\label{transferm}
\ee
where in the last line we recall that $Z_\rf$ has vanishing fermionic components. The expression in equation~\eqref{NMHVdualco} is therefore just the dual superconformal invariant $[\rf,i\!-\!1,i,j\!-\!1,j]$, for the specific choice $Z_\rf = (0,\iota^{A'},0)$ of reference twistor. The components of $Z_\rf$ may be changed to arbitrary values by applying a dual superconformal transformation. Since the sum~\eqref{NMHVtree1} is both independent of the spinor $|\iota]$ and is a dual superconformal invariant,  it must in fact be independent of the entire reference supertwistor $Z_\rf$. Thus we have proved that our rule yields a correct expression for $M_{\rm NMHV}$ on momentum twistor space.

Observe that the BCFW form~\cite{Drummond:2008vq} of the NMHV tree
\be
	M_{\rm NMHV} = \sum_{1< i<j< n} [n,i\!-\!1,i,j\!-\!1,j] \quad\left(= \sum\limits_{1< i<j< n} R_{n;ij}\right)
\label{BCFWNMHV}
\ee
follows immediately from the MHV diagram form~\eqref{NMHVtree1} upon setting $Z_\rf = Z_n$ and recalling that $[\ ,\,,\,,\,,\,]$ vanishes if any two entries are the same. (This equality may in fact be proven directly, and without fixing $Z_\rf$, by expanding each BCFW term using the six term
identity
\begin{multline}
	[i\!-\!1,i,j\!-\!1,j,n] + [i,j\!-\!1,j,n,\rf] + [j\!-\!1,j,n,\rf,i\!-\!1] \\
	+[j,n,\rf,i\!-\!1,i] + [n,\rf,i\!-\!1,i,j\!-\!1] + [\rf,i\!-\!1,i,j\!-\!1,j] =0
\end{multline}
and noticing that all terms depending on $Z_n$ cancel in pairs, leaving
precisely the MHV diagram terms.) We shall return to this observation in a future publication.

%%%%%%%%%%%%%%%%%%%%%%%%%%%%%%%%%%%%%%%%%%%%%%%%%%%
%%%%%%%%%%%%%%%%%%%%%%%%%%%%%%%%%%%%%%%%%%%%%%%%%%

\subsection{N$^2$MHV tree amplitudes}
\label{sec:NNMHVtree}

\begin{figure}[t]
	\centering
	\includegraphics[height=23mm]{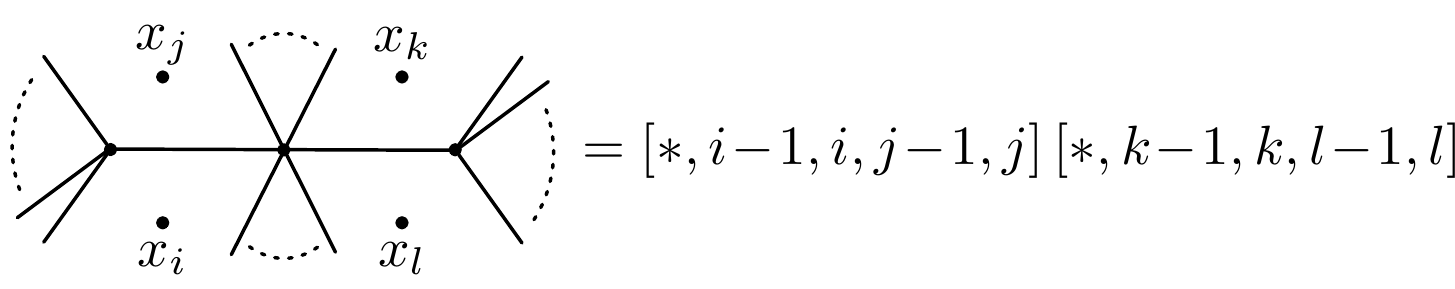}
	\caption{An MHV diagram contributing to the tree-level N$^2$MHV amplitude, together with its associated 
			product of dual superconformal invariants}
	\label{fig:CSWNNMHV}
\end{figure}

We now consider MHV diagrams that contribute to the tree-level N$^2$MHV amplitude. These have three vertices connected by two propagators as shown in figure \ref{fig:CSWNNMHV}. We take the convention that the first external leg attached to the leftmost vertex in a clockwise cyclic ordering is always $i$. Notice that we must have $i<j-1$ and $k <l-1$ for the left and right vertices to have at least three legs.

For the generic case $l\neq i$ and $k\neq j$, there are no adjacent propagators, so our rules assign the product
\be
	[\,\rf\,,i\!-\!1,i,j\!-\!1,j]\,[\,\rf\,,k\!-\!1,k,l\!-\!1,l]\,.
\ee
The proof that our rule is correct here follows essentially identically to the NMHV case presented above. Once the overall tree-level MHV superamplitude has been stripped off and the fermionic integrations performed we are are left with the product of two R-invariants as claimed.

\begin{figure}[t]
	\centering
	\includegraphics[height=23mm]{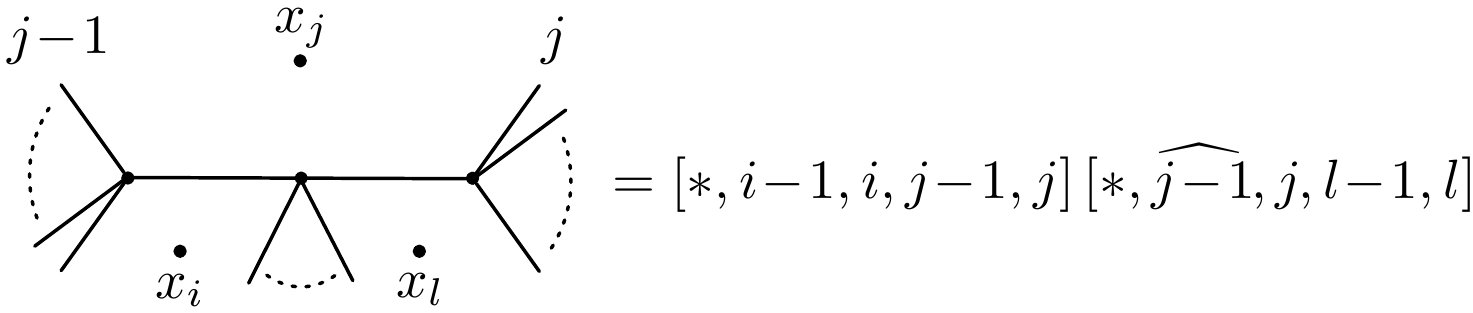}
	\caption{MHV diagrams occurring in the tree-level $\Nsq$ amplitude for which the two propagators are adjacent.
	Notice that $j\!-\!1$ is unshifted in the first R invariant, because the external leg $j\!-\!1$ is attached to the
	propagator on the left. Notice also that, with our choice of orientation, the external leg $j$ is never shifted.}
	\label{fig:NNMHVboundary}
\end{figure}

In the boundary cases where either $l=i$ or $j=k$, the two propagators are adjacent in the cyclic ordering of the middle MHV vertex, so we instead obtain
\be
	[\,\rf\,,\widehat{i\!-\!1},i,j\!-\!1,j]\,[\,\rf\,, k\!-\!1,k,i\!-\!1,i]
\ee
in the case $l=i$, where $\widehat{i\!-\!1} = (i\!-\!1,i)\cap(\,\rf\,,k\!-\!1,k)$, or else
\be
	[\,\rf\,,i\!-\!1,i,j\!-\!1,j]\,[\,\rf\,,\widehat{j\!-\!1},j,l\!-\!1,l]
\ee
in the case $k=j$, where $\widehat{j\!-\!1} = (j\!-\!1,j)\cap(\rf,i\!-\!1,i)$ (see figure~\ref{fig:NNMHVboundary}). We never have $l=i$ and $k=j$ simultaneously, because the middle vertex must have at least three legs. Combining the terms gives
\be
	M_{{\rm N}^2{\rm MHV}}= \sum\,[\,\rf\,,\widehat{i\!-\!1},i,j\!-\!1,j]\,[\,\rf\,,\widehat{k\!-\!1},k,l\!-\!1,l]
\label{NNMHVtree}
\ee
where the sum is over the range $1\leq i< j\leq k < l\leq i+n$ (understood mod $n$) and where
\be
	\widehat{i\!-\!1}=
		\begin{cases}
			(i\!-\!1,i)\cap(\,\rf\,,k\!-\!1,k) & \hbox{if } l= i\\
			\,i\!-\!1 & \hbox{otherwise,} 
		\end{cases}
\ee
and
\be
	\widehat{k\!-\!1}=
		\begin{cases}
			(k\!-\!1,k)\cap(\,\rf\,,i\!-\!1,i) &  \hbox{if } k=j\\
			\,k\!-\!1 & \hbox{otherwise.}
		\end{cases}
\ee

We wish to show that the shifted terms in our rule correctly account for such `boundary' diagrams with adjacent propagators.  To this end, consider the N$^2$MHV diagram with $j=k$ shown in figure~\ref{fig:NNMHVboundary}. There are two  propagators adjacent to the region $x_j$, separating it from $x_i$ and $x_k$, respectively.  Once again pulling out the overall MHV superamplitude, the usual momentum space MHV rules lead to the expression  
\be
\label{N2MHV-adj}
	\frac{\la i\!-\!1\, i\ra\la j\!-\!1\, j\ra\,\delta^{0|4}([\iota|x_{ij}|\theta_{ji}\ra) } 
	{x_{ij}^2 \,\la i\!-\!1\,\lambda\ra\,\la\lambda\,i\ra\,\la j\!-\!1\,\lambda\ra}
	\,\times\,\frac{1}{\la\lb\,\lb'\ra}\,\times\,
	\frac{\la l\!-\!1\,l\ra\,\delta^{0|4}( [\iota|x_{jl}|\theta_{lj}\ra)}	{x_{jl}^2\,\la\lb'\,j\ra\,\la\l\!-\!1\,\lb'\ra\,\la\lb'\,l\ra}\,,
\ee
where $\la\lb| \equiv [\iota|x_{ij}$ and $\la\lb'|\equiv [\iota|x_{jl}$ are the CSW prescriptions for the spinors associated to the propagators. To compare this to our expression $[\rf,i\!-\!1,i,j\!-\!1,j]\,[\rf,\widehat{j\!-\!1},j,l\!-\!1,l]$ note first that
\be
\begin{aligned}
	&{[\rf,\widehat{j\!-\!1},j,l\!-\!1,l]}\equiv\\
	&\ \ \frac{\delta^{0|4}(\la\rf,\widehat{j\!-\!1},j,l\!-\!1\ra\chi_l + \hbox{cyclic})}
	{\la \rf,\widehat{j\!-\!1},j,l\!-\!1\ra\,\la\widehat{j\!-\!1},j,l\!-\!1,l\ra \,\la j,l\!-\!1,l,\rf \ra\,\la l\!-\!1,l,\rf,\widehat{j\!-\!1}\ra\,
	\la l,\rf,\widehat{j\!-\!1},j\ra}\\
	&=\frac{\la\rf,i\!-\!1,i,j\ra\ \delta^{0|4}(\la\rf,j\!-\!1,j,l\!-\!1\ra\chi_l + \hbox{cyclic})}
	{\la \rf,j\!-\!1,j,l\!-\!1\ra\,\la j\!-\!1,j,l\!-\!1,l\ra\,\la j,l\!-\!1,l,\rf\ra\,\la l\!-\!1,l,\rf,\widehat{j\!-\!1}\ra\,\la l,\rf,j\!-\!1,j\ra}\,
\end{aligned}
\label{shifted-invt} 
\ee
where in going to the second line, we replaced $\widehat{Z_{j-1}}$ by $Z_{j-1}$ in all terms where it is skewed with $Z_j$, at the expense of a factor of $\la\rf,i\!-\!1,i,j\ra$ (four from the numerator and three from the denominator).
The remaining expression involving the shifted twistor is
\be
	\la l\!-\!1,l,\rf,\widehat{j\!-\!1}\ra = \la\rf,l\!-\!1,l,\left[j\!-\!1\ra,\la j\right],i\!-\!1,i,\rf\ra
	\propto [\iota|x_{li}x_{ij}|\iota] = \la\lb'\,\lb\ra
\ee
where we have used the translation~\eqref{transbos}-\eqref{transferm} between momentum twistors and momentum space. Translating~\eqref{shifted-invt} to momentum space thus gives
\be
	\la\lb\,j\ra\,\times\,\frac{\la l\!-\!1\,l\ra \,\delta^{0|4}([\iota|x_{jl}|\theta_{lj}\ra)}
	{x_{jl}^2\,\la\lb'j\ra\, \la l\!-\!1\,\lb'\ra\,\la\lb'\,l\ra\,\la\lb\,\lb'\ra\,}\,.
\ee
It is easy to check that the unwanted numerator factor $\la\lb\,j\ra=[\iota|x_{ij}|j\ra=\la\lb\,j\ra$ (coming from the momentum twistor numerator $\la\rf,i\!-\!1,i,j\ra$) cancels with a similar factor in the denominator of $[\rf,i\!-\!1,i,j\!-\!1,j]$. The remaining factors from $[\rf,i\!-\!1,i,j\!-\!1,j]$ provide the remaining pieces of~\eqref{N2MHV-adj}, as required.

Summing over the possible diagrams, we find that the expression~\eqref{NNMHVtree} for $M_{{\rm N}^2{\rm MHV}}$ on momentum twistor space given by our rules is indeed correct. Once again, although our assignment of R invariants to MHV diagrams only agrees with the standard momentum space expressions  term-by-term if we choose $Z_\rf$ to have the special form $(0,\iota^{A'},0)$, the sum~\eqref{NNMHVtree} is independent of $\iota$ and dual superconformally invariant. It is thus independent of the complete supertwistor $Z_\rf$.

\medskip

As we remarked above, the rule for assigning R invariants to propagators requires that we choose an
orientation of the planar diagram (here taken to be the clockwise orientation). An external leg is only shifted if it \emph{precedes} the given propagator in this ordering, so that the external leg $j$ in figure~\ref{fig:NNMHVboundary} is never shifted. With the opposite choice orientation, for the same diagram, $j\!-\!1$ would remain unshifted whilst we would instead replace
$$
	[\,\rf\,,i\!-\!1,i,j\!-\!1,j] \longrightarrow [\,\rf\,,i\!-\!1,i,j\!-\!1,\widehat{\jmath}\,]
$$
in the first R invariant, with $\widehat{\jmath}\equiv (j\!-\!1,j)\cap(\rf,l\!-\!1,l)$. That the two choices are equal follows from the identity
\be
	[\,\rf\,,i\!-\!1,i,j\!-\!1,j]\,[\,\rf\,,\widehat{j\!-\!1},j,l\!-\!1,l] = [\,\rf\,,i\!-\!1,i,j\!-\!1,\widehat{\jmath}\,]\,[\,\rf\,,j\!-\!1,j,l\!-\!1,l]
\label{N2MHV-exps}
\ee
which is a consequence of the orientation reversal symmetry of the diagram.

We also remark that although the product of two R invariants is always manifestly dual conformal invariant, dual {\em super}conformal invariance of an R invariant that involves a shifted twistor only follows on the support of the fermionic delta function in the other R invariant. Thus, whilst $[\rf,\widehat{j\!-\!1},j,l\!-\!1,l]$ is dual conformally invariant (up to a choice of $Z_\rf$), it is not dual superconformal unless multiplied by $[\,\rf\,,i\!-\!1,i,j\!-\!1,j]$. Geometrically, this is because $\widehat{Z_{j-1}}$ was defined as the intersection of the line $(j\!-\!1,j)$ with the plane $(\rf , i\!-\!1,i)$. However,  \emph{super}twistor space has dimension 8, so that in general lines and planes do not intersect -- they can `miss' in the fermionic directions.  On the support of the fermionic $\delta$-function in $[\,\rf\,,i\!-\!1,i,j\!-\!1,j]$ however, $Z_\rf$, $Z_{i-1}$, $Z_i$, $Z_{j-1}$ and $Z_j$ span a four-dimensional subspace of the full \emph{super}twistor space. Within this subspace, the required intersection will always take place.

%%%%%%%%%%%%%%%%%%%%%%%%%%%%%%%%%%%%%%%%%%%%%%%%%%%
%%%%%%%%%%%%%%%%%%%%%%%%%%%%%%%%%%%%%%%%%%%%%%%%%%%

\subsection{A diagram for the N$^3$MHV tree}
\label{sec:N3MHVtree}

\begin{figure}[t]
	\centering
	\includegraphics[height=32mm]{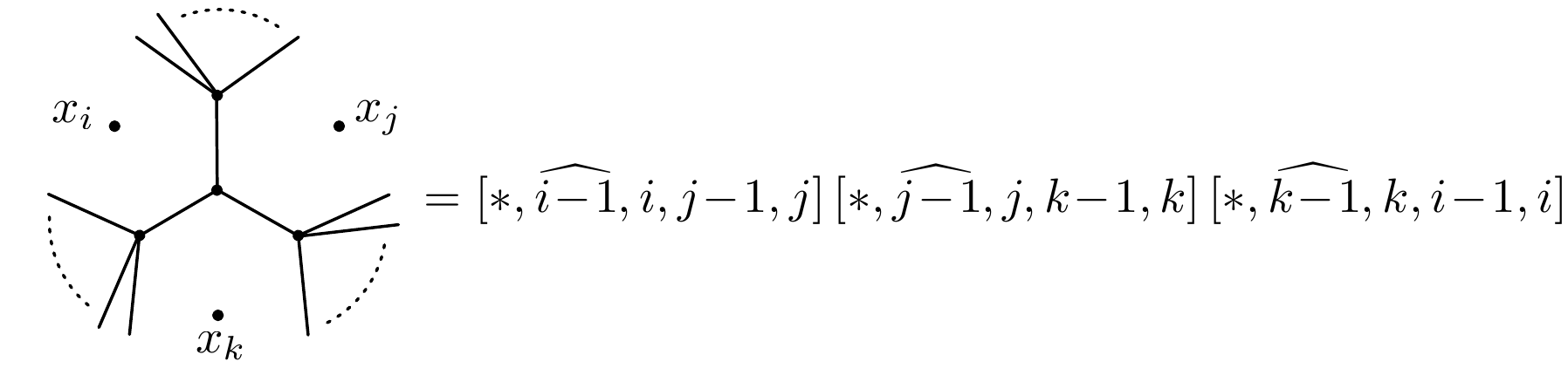}
	\caption{An MHV diagram contributing to the N$^3$MHV tree amplitude. This diagram illustrates the use
	of concatenated shifts.}
	\label{fig:N3MHVtree}
\end{figure}

The final tree example we consider is a particular diagram that contributes to the $n$-particle N$^3$MHV tree, illustrated in figure~\ref{fig:N3MHVtree}. The MHV rules thus assign a factor
\be
	[\,\rf\,,\widehat{i\!-\!1},i,j\!-\!1,j]\,[\,\rf\,,\widehat{j\!-\!1},j,k\!-\!1,k]\,[\,\rf\,,\widehat{k\!-\!1},k,i\!-\!1,i]
\ee
to this diagram. 

In this diagram we see a new phenomenon: that a propagator can be adjacent to others on both sides.  In fact in this case, only one twistor is shifted in the R-invariant for each propagator, although one can envisage diagrams in which two need to be shifted.  The propagator separating $x_i$ from $x_j$ is connected to the vertex on which external leg $j\!-\!1$ ends, but is not directly attached to the vertex on which $i\!-\!1$ ends. Thus, in the first R-invariant $j\!-\!1$ is unshifted whereas $i\!-\!1$ is shifted. The shift is determined by the region $x_k$ associated to the $x_i$, $x_k$ propagator -- the propagator that precedes the $x_i$, $x_j$ propagator in the direction of $i\!-\!1$. Altogether, the appropriate 
shifts for all the variables are
\be
	\widehat{i\!-\!1} = (i\!-\!1,i)\cap(\,\rf\,,k\!-\!1,k)\,\quad
	\widehat{j\!-\!1} = (j\!-\!1,j)\cap(\,\rf\,,i\!-\!1,i)\,\quad
	\widehat{k\!-\!1} = (k\!-\!1,k)\cap(\,\rf\,,j\!-\!1,j)
\ee
according to the momentum twistor MHV rule.

%%%%%%%%%%%%%%%%%%%%%%%%%%%%%%%%%%%%%%%%%%%%%%%%%%%
%%%%%%%%%%%%%%%%%%%%%%%%%%%%%%%%%%%%%%%%%%%%%%%%%%

\subsection{General trees}
\label{sec:gentrees}

We now show that a general tree-level MHV diagram may be written by assigning an R invariant to each propagator in accordance with the rules given at the beginning of the section. Our rule involves orienting the diagram in a clockwise sense; there is an equivalent, though differently represented rule for the anticlockwise orientation.

To reformulate a general tree amplitude in momentum twistor space, we must divide through by the overall MHV tree and perform the fermionic integrals associated to each propagator using the remaining fermionic $\delta$-functions in the MHV vertices. The general procedure is as follows. Multiplying through by $\prod \la i-1\, i\ra$ is straightforward. On the support of Grassmann $\delta$-functions at all the other vertices, the $\delta^{0|8}$-function at any particular vertex can be made to depend purely on the external states. It then becomes the overall factor $\delta^{0|8}(\sum_{i=1}^n |i\ra \eta_i)$ and may be removed.  Any (connected) tree diagram obeys
\be
	V-1=P
\ee
so there is a Grassmann $\delta^{0|8}$-function remaining for each of the $P$ propagators. In particular, on their mutual support, we can reassemble them into a product with one for each propagator so that for the propagator separating region $x_i$ from $x_j$, this $\delta$-function is $\delta^{0|8}(\theta_{ij} + |\lambda\ra\eta)$. Here,  $|\lambda\ra=x_{ij}|\iota]$ is the CSW prescription for the propagator momentum and $\eta$ is the fermionic variable of the propagator, associated with the sum over the $\cN=4$ supermultiplet passing along this propagator. Performing the integration over $\eta$ gives
\be
	\int\rd^4\eta\ \delta^{0|8}(\theta_{ij} + |\lambda\ra\eta = \delta^{0|4}(\la\lb|\theta_{ij}\ra)
\ee
and these remaining $4P$ fermionic $\delta$-functions will form the Grassmann parts of the $P$ R invariants.

For every propagator that is in between two external lines in the cyclic ordering of each of its terminating vertices, the calculation is identical to the NMHV case considered above and gives simply $[\,\rf\,,i\!-\!1,i,j\!-\!1,j]$.  However, the diagram may contain arbitrarily long sequences of propagators that are all adjacent to the same region, say $x_i$. Furthermore, two such sequences may  `back onto' each other, as in the N$^3$MHV diagram considered above. Diagrams of this type are in fact generic for anti-MHV amplitudes. Our task in what follows is to prove that the arguments of the corresponding R invariants are always shifted in accordance with the rule given above.

\begin{figure}[t]
\centering
\includegraphics[height=25mm]{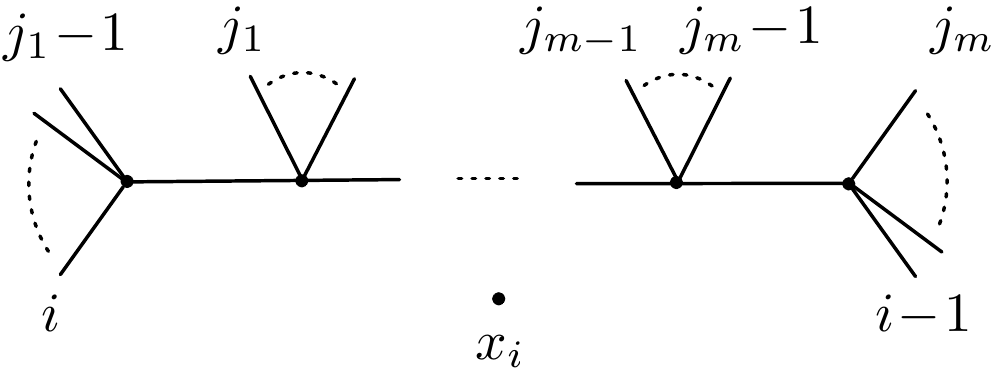}
\caption{A sequence of propagators that are all adjacent to the region $x_i$.}
\label{fig:AdjacentPropagators}
\end{figure}

We first consider a sequence of $m$ propagators that are all adjacent to the region $x_i$, {\it i.e.}, are all in between external legs $i\!-\!1$ and $i$ (see figure \ref{fig:AdjacentPropagators}). Let us assume for the moment
that this sequence does not back onto any other such sequence.  Then the $r^{\rm th}$ such propagator will
has adjacent external legs $j_r\!-\!1$ and $j_r$.  Reading from the right of the figure,  the contribution to the MHV diagram from this sequence of propagators is 
\be
\begin{aligned}
\label{seq} 
	&\frac{\la i\!-\!1\, i\ra\; \la j_1\!-\!1\, j_1\ra\; \delta^{0|4}([\iota|x_{ij_1}|\theta_{j_1i}\ra)}
	{x_{ij_1}^2 \;[\iota|x_{ij_1}|j_1\!-\!1\ra \;[\iota|x_{ij_1}|j_1\ra\;  [\iota | x_{ij_1}|i\ra}\\
	&\qquad\qquad\times \left(\prod_{r=2}^{m} 
	\frac{\la j_r\!-\!1\, j_r\ra\;\delta^{0|4}([\iota|x_{ij_r}|\theta_{j_ri}\ra) } {x_{ij_r}^2 \; [\iota |x_{ij_r}|j_r\!-\!1\ra \;
	[\iota |x_{ij_r}|j_r\ra \; [\iota |x_{j_r i}\,x_{ij_{r+1}}|\iota]}\right)
	\, \times\, \frac{1}{[\iota|x_{ij_m}|i\!-\!1\ra}\\
	&=\   \frac{\la i\!-\!1\, i\ra\; \la j_1\!-\!1\, j_1\ra\; \delta^{0|4}([\iota|x_{ij_1}|\theta_{j_1 i}\ra)}
	{x_{ij_1}^2 \;[\iota|x_{ij_1}|j_1\!-\!1\ra \;[\iota|x_{ij_1}|j_1\ra\;[\iota|x_{ij_1}|i\ra\;[\iota|x_{ij_1}|i\!-\!1\ra} \\
	&\qquad\qquad\times\ \prod_{r=2}^{m} 
	\left(\frac{[\iota|x_{ij_{r-1}}|i\!-\!1\ra}{[\iota|x_{ij_r}|i\!-\!1\ra}\times
	\frac{\la j_r-1\, j_r\ra\; \delta^{0|4}([\iota|x_{ij_r}|\theta_{j_ri}\ra) }
	{x_{ij_r}^2 \; [\iota |x_{ij_r}|j_r\!-\!1\ra \;  [\iota |x_{ij_r}|j_r\ra \;  [\iota |x_{j_r i}\,x_{ij_{r+1}}|\iota]}\right)\, .
\end{aligned}
\ee
In the second version we have included a ratio in the product that cancels one term with the next to leave
just one overall factor of $[\iota|x_{ij_m}|i\!-\!1\ra$ in the denominator. This product can now straightforwardly be identified with
\be
	\prod_{r=1}^{m}\, [\,\rf\,,(\widehat{i\!-\!1})_r, i,j_r\!-\!1,j_r]
	\ =\ 
	\prod_{r=1}^m\, [\,\rf\,,i\!-\!1,\widehat{\imath_r},j_r\!-\!1,j_r]\,,
\label{seq-inv}
\ee
where 
\be
	(\widehat{i\!-\!1})_r = 
		\begin{cases}
			(i\!-\!1,i)\cap(\,\rf\,,j_{r+1}\!-\!1,j_{r+1}) & \hbox{for } r\neq m\\
			\ i\!-\!1 & \hbox{for } r = m
		\end{cases}
\ee
and
\be
	\hat{\imath}_r  = 
		\begin{cases}
			(i\!-\!1,i)\cap(\,\rf\,,j_{r-1}\!-\!1,j_{r-1}) & \hbox{for } r\neq 1 \\
			\ i & \hbox{for } r=1\,.
		\end{cases}
\ee
The equality of the two expressions in~\eqref{seq-inv} follows from the performing the same calculation with the orientation reversed.

Finally we need to check that the given prescription holds even when, for some values of $r$, we have $j_{r-1}=j_r$ or a longer sequence of adjacent propagators on the other side.  The key point is that each R-invariant has 5 denominator factors.  In our prescription it is always the case that one of them gives the actual physical propagator $1/x_{ij}^2$, and two of them correspond to spinor products in the Parke-Taylor denominator of the MHV vertex that are associated with the region on one side $x_i$, and two with the region on the other side $x_j$.  It is easily seen that the factors associated to region $x_i$ operate independently of those that are adjacent to region $x_{j}$.  Thus we can handle this situation by applying separate shifts to the corresponding momentum twistors  on both sides of the propagator ($i\!-\!1$ and $j_r\!-\!1$) according to the given prescription.

We emphasize that the overall orientation is irrelevant; the orientation of  each sequence of adjacent propagators can be chosen independently from each other. Each choice merely corresponds to a different representation of the same underlying formula.

%%%%%%%%%%%%%%%%%%%%%%%%%%%%%%%%%%%%%%%%%%%%%%%%%%%
%%%%%%%%%%%%%%%%%%%%%%%%%%%%%%%%%%%%%%%%%%%%%%%%%%

%%%%%%%%%%%%%%%%%%%%%%%%%%%%%%%%%%%%%%%%%%%%%%%%%%%
%%%%%%%%%%%%%%%%%%%%%%%%%%%%%%%%%%%%%%%%%%%%%%%%%%

%%%%%
%%%Loop Amplitudes
%%%%%

\section{MHV loop diagrams in momentum twistor space}
\label{sec:MHVloops}

In this section we consider using MHV diagrams for loop amplitudes in $\cN=4$ SYM.  At one loop, the Feynman tree theorem has been used~\cite{Brandhuber:2005kd} to prove the equivalence of the MHV formalism with conventional methods. However, beyond the 1-loop MHV amplitude computed in~\cite{Brandhuber:2004yw}, we are not aware 
of any  explicit calculations of amplitudes or integrands in $\cN=4$ SYM using the MHV formalism. In large part this is because in momentum space, the formal expression of a sum over products of MHV vertices with the CSW prescription for propagators is more complicated than competing methods such as generalized unitarity. We will see that working in momentum twistor space considerably simplifies the formul\ae.

Our first task in this section is to extend the tree-level rule for assigning an R-invariant to each propagator in a loop diagram. The main new feature is that the R-invariants can now depend on momentum twistors associated to the loop variables. In a planar diagram, all the propagators that form a given loop must be adjacent in the cyclic ordering of the MHV vertices around the edge of that loop. Our tree-level rule for shifting the arguments of these R-invariants must be extended to cope with this situation.

Using our loop rule, MHV diagrams quickly present the integrand\footnote{
	As emphasised in~\cite{Arkani-Hamed:2010kv}, `the integrand' of a planar diagram is meaningful
	only when working in terms of region momenta (or momentum twistors) where momentum conservation
	is manifest. This removes the translational freedom in the loop momentum variable, allowing us
	to sensibly add together loop integrands from different diagrams.
	} 
of a planar $\cN=4$ SYM loop amplitude in terms of a sum of products of R invariants.  In section~\ref{sec:loopintegral} we go on to show that these expressions can be easily manipulated into a form that is suitable for integration~\cite{Drummond:2010mb}.

%%%%%%%%%%%%%%%%%%%%%%%%%%%%%%%%%%%%%%%%%%%%%%%%%%%
%%%%%%%%%%%%%%%%%%%%%%%%%%%%%%%%%%%%%%%%%%%%%%%%%%

\subsection{Momentum twistor MHV rules at one loop}
\label{sec:OneLoopRule}

One loop amplitudes have an integral over a single `internal' region momentum, corresponding to an arbitrary line in momentum twistor space. We suppose that X is the line $(A,B)$ through some points $Z_A$ and $Z_B$. At one loop, the momentum twistor MHV rules are then
\begin{itemize}
	\item To each propagator separating the external regions $x_i$ and $x_j$, assign a factor of 
		$[\,\rf\,,\widehat{i\!-\!1},i,\widehat{j\!-\!1},j]$ exactly as at tree-level.
	
	\item To each propagator separating an external region $x_i$ from the internal region $x_{AB}$, assign a factor of
		$[\,\rf\,,\widehat{i\!-\!1},i,A,\hat{B}]$.
\end{itemize}
In these loop R invariants, the shifted external leg $\widehat{i\!-\!1}$ is determined separately for each propagator in exactly they same way as at tree-level. The shifted loop variable $\hat B$ is also determined separately for each propagator according to the rule:
\begin{itemize}
	\item For each propagator in the loop, $\hat{B} = (AB)\cap(\,\rf\,,k\!-\!1,k)$ where $k\!-\!1$ and $k$ are the
		external legs associated to the preceding loop propagator in the cyclic ordering.
\end{itemize}
As at tree-level, this rule assumes that we have picked an orientation of the planar diagram, which below we take to be clockwise. Again there is an analogous rule for the anticlockwise orientation.

Following this rule leads to a representation of a 1-loop MHV diagram contributing to the planar N$^k$MHV amplitude as a product of $P$ R-invariants, of total Grassmann degree $4P$, and depending on the \emph{super}twistors $Z_A$ and $Z_B$ associated with the loop. Integrating out the fermionic components of $Z_A$ and $Z_B$ leaves us with total Grassmann degree $4(P-2)$, or $4(P-2L)$ at arbitrary loop order.  From the momentum space MHV rules, we expect to have Grasssmann degree $8V$ from the MHV vertices, less $4P$ from the sum over the supermultiplets running along each propagator, less a further 8 for overall super-momentum conservation.  For any (connected) loop diagram
\be
	P-V=L-1
\ee
so $8V-4P-8 = 4(P-2L)$, agreeing with Grassmann degree provided by the momentum twistor MHV rule.

%%%%%%%%%%%%%%%%%%%%%%%%%%%%%%%%%%%%%%%%%%%%%%%%%%%
%%%%%%%%%%%%%%%%%%%%%%%%%%%%%%%%%%%%%%%%%%%%%%%%%%

\subsection{MHV 1-loop}
\label{sec:MHV1loop}

\begin{figure}[t]
	\centering
	\includegraphics[height=25mm]{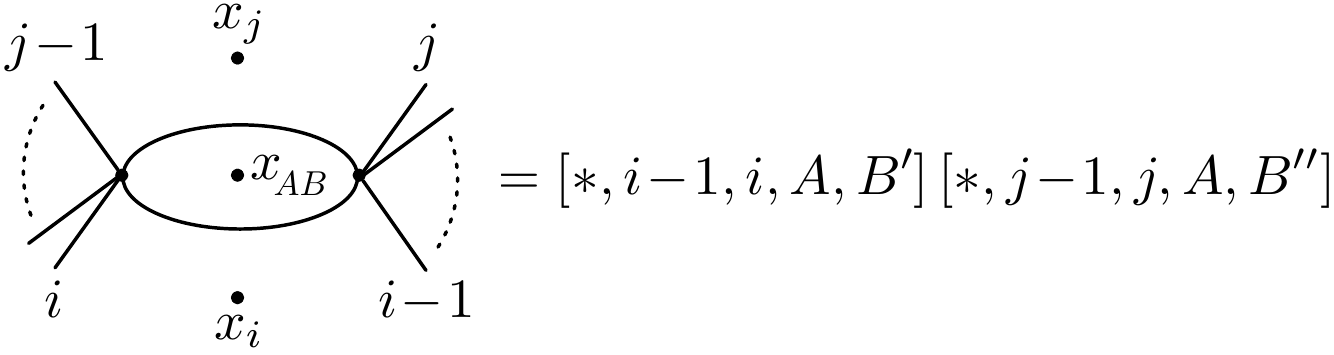}
	\caption{Every MHV diagram that contributes to the 1-loop planar MHV amplitude has the topology of a bubble, for
			some choice of $i$ and $j$.}
	\label{fig:1loopMHV}
\end{figure}

We begin by illustrating the use of this rule with the simple example of the 1-loop MHV planar integrand. Every MHV diagram that contributes to this amplitude has the form shown in figure~\ref{fig:1loopMHV}. Note that all the external lines are attached to a vertex on which both propagators end. Therefore, according to the above rules, the lower propagator should be assigned the R invariant $[\,\rf\,,i\!-\!1,i,A,B']$ with $B' = (A,B)\cap(\,\rf\,,j\!-\!1,j)$ while the upper propagator is likewise assigned the R invariant $[\,\rf\,,j\!-\!1,j,A,B'']$ with $B'' = (A,B)\cap(\,\rf\,,i\!-\!1,i)$. Multiplying the two and summing over the possible diagrams, we find that the 1-loop planar MHV amplitude may be written in momentum twistor space as
\be
	M_{\rm MHV}^{(1)} = \int{\rm D}^{3|4}Z_A\wedge{\rm D}^{3|4}Z_B \sum_{i<j} \,[\,\rf\,,i\!-\!1,i,A,B']\,[\,\rf\,,j\!-\!1,j,A,B'']\,.
\label{MHV1loop}
\ee
As discussed above, the integration over the fermionic parts of the auxiliary twistors $Z_A$ and $Z_B$ is necessary to provide the correct Grassmann degree. As will be discussed further in section~\ref{sec:loopintegral}, the integration over the bosonic parts of these twistors can be interpreted as the loop integral itself. Note that the summation range $i<j$ here allows $i=j\!-\!1$, corresponding to a three-point MHV vertex on one end of the loop. 

Let us check that~\eqref{MHV1loop} is correct. In momentum space, once we pull out the overall momentum conserving $\delta$-function, the standard MHV rules associate the diagram in figure~\ref{fig:1loopMHV} with the expression~\cite{Brandhuber:2004yw,Bena:2004xu}
\be
	\int \rd^4x_{AB}\left(\int \rd^4\eta_i \,\rd^4\eta_j \  
	\frac{A_{\rm MHV}(\ell_i,i,\ldots,j-1,\ell_j)A_{\rm MHV}(\ell_j, j \ldots,i-1,\ell_i)}{(x_{AB}-x_i)^2(x_{AB}-x_j)^2}
	\right)\,,
\ee
where
\be
 		|\ell_i\ra = (x_{AB}-x_i)|\iota ] \qquad \hbox{and} \qquad |\ell_j\ra = (x_{AB}-x_j)|\iota ]
\label{CSWloopspinors}
\ee
are the CSW prescription for the spinors associated to the momentum in the lower and upper propagators, respectively. (Again, $|\iota]$ is a reference spinor.) Pulling out an $n$-particle Parke-Taylor denominator and performing the fermionic loop integrals, the remaining integrand is
\be
	\frac{1}{(x_{AB}-x_i)^2 (x_{AB}-x_j)^2}
	\frac{\la\ell_i\,\ell_j\ra^2}{\la\ell_i\,i\!-\!1\ra\,\la\ell_i\,i\ra\, \la\ell_j\,j\!-\!1\ra\, \la\ell_j\,j\ra}\, .
\ee

\begin{figure}[ht]
	\centering
	\includegraphics[height=35mm]{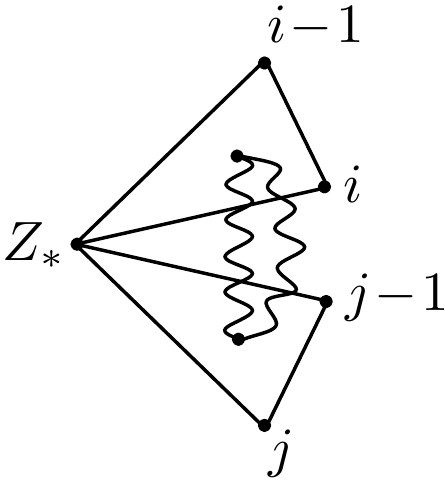}
	\caption{A representation of the MHV diagram in the previous figure after integrating out $\chi_A$
			and $\chi_b$. Each solid line represents a factor of $\la A,B,\,\cdot\,,\,\cdot\,\ra$, with the remaining
			two entries determined by the ends of the line in the figure. The two wavy lines represents the (squared) 
			numerator factors. An analogous figure for the MHV integrand produced by the BCFW recursion
			relation appears in~\cite{Arkani-Hamed:2010kv}.}
\label{fig:kermit}
\end{figure}

To translate this expression to momentum twistor space, we again promote the reference spinor to a reference momentum twistor $Z_\rf = (0,\iota^{A'},0)$. We then find 
\be
\label{MHV1loopmomtw}
		\frac{1}{\la A,B,i\!-\!1,i \ra\,\la A,B,j\!-\!1,j \ra}\times
		\frac{\left(\la\,*\,,i\!-\!1,i,\left[A\ra,\la B\right],j\!-\!1,j,\,\rf\,\ra\right)^2}
		{\la A,B,i\!-\!1,\,\rf\,\ra\,\la A,B,i,\,\rf\,\ra\,\la A,B, j\!-\!1,\,\rf\,\ra\,\la A,B,j,\,\rf\,\ra}\,.
\end{equation}
The numerator of this expression comes from $\la\ell_i\,\ell_j\ra^2$ and may be understood as the square of ${\rm X}^{\alpha\beta}{\rm Y}_{\alpha\beta}$, where X is the line $(A,B)$ and ${\rm Y} = (\,\rf\,,i\!-\!1,i)\cap(\,\rf\,,j\!-\!1,j)$. Following~\cite{Arkani-Hamed:2010kv}, we can summarise the ratio~\eqref{MHV1loopmomtw} with the diagram in figure~\ref{fig:kermit}.

To compare~\eqref{MHV1loopmomtw} to the product of R invariants given by our rules for the same diagram, we must integrate out the fermionic components of the loop supertwistors $Z_A$ and $Z_B$ in~\eqref{MHV1loop}.  This is straightforward:  first notice that
\be
	\delta^{0|4}\!\left(\chi_A\la B',i\!-\!1,i,\,\rf\,\ra+\hbox{cyclic}\right) 
	=\la \,\rf\,j\!-\!1,j,A\ra^4\,\delta^{0|4}\!\left(\chi_A\la B,i\!-\!1,i,\,\rf\,\ra+\hbox{cyclic}\right)
\ee
and similarly for the second R invariant; then integration over the Grassmann $\delta$-functions gives a factor
\be
	\la\,\rf\,,i\!-\!1,i,A\ra^4\la\,\rf\,,j\!-\!1,j,A\ra^4\, \left(\la\,*\,,i\!-\!1,i,\left[A\ra,\la B\right],j\!-\!1,j,\,\rf\,\ra\right)^4
\ee
which combines with the remaining factors in the denominator of the product of the two R-invariants to give exactly~\eqref{MHV1loopmomtw}. 

This proves that the loop integrand~\eqref{MHV1loop} provided by the MHV rules in momentum twistor space agrees with the standard MHV rules expression for the loop integrand when the reference supertwistor is $Z_\rf = (0,\iota^{A'},0)$. That the momentum space MHV rules agree with standard Feynman rules for the 1-loop amplitude was proved in~\cite{Brandhuber:2005kd}. The loop \emph{integrand} is dual superconformally invariant~\cite{Arkani-Hamed:2010kv}, so as at tree level we are free to use the momentum twistor MHV rules with a completely arbitrary choice of reference twistor. 

The sum over MHV diagrams allows $j=i\!+\!1$, whose corresponding diagrams involve a three-point MHV vertex with external particle $i$. For these terms,  the above formul\ae\ simplify using the identity
\be		 
	\la\,\rf\,,i\!-\!1,i,A\ra\,\la B,i,i\!+\!1,\,\rf\,\ra + \la\,\rf\,,i\!-\!1,i,B\ra\,\la A,i\!+\!1,i,\,\rf\,\ra
	+ \la\,\rf\,,i\!-\!1,i,i\!+\!1\ra\,\la A,B,i,\,\rf\,\ra=0\, .
\ee
The numerator factor $\la A,B,i,\,\rf\,\ra^2$ cancels two spurious propagators  resulting in a box integrand with one spurious and three physical propagators.

\medskip

Incidentally, note again that if we set $Z_\rf$ to be the equal to the external twistor $Z_n$, then~\eqref{MHV1loop} becomes
\be
\begin{aligned}
	M_{\rm MHV}^{(1)} &= \int{\rm D}^{3|4}Z_A\wedge{\rm D}^{3|4}Z_B 
	\sum_{1<i<j< n} \,[n,i\!-\!1,i,A,B']\,[n,j\!-\!1,j,A,B'']\\
	&=\int{\rm D}^3Z_A\wedge{\rm D}^3Z_B \sum_{1<i<j<n}\left(\frac{1}{\la A,B,i\!-\!1,i \ra\,\la A,B,j\!-\!1,j \ra}\right.\\
	&\hspace{4cm}\left.\times\ \frac{\left(\la n,i\!-\!1,i,\left[A\ra,\la B\right],j\!-\!1,j,n\ra\right)^2}
		{\la A,B,i\!-\!1,n\ra\,\la A,B,i,n\ra\,\la A,B, j\!-\!1,n\ra\,\la A,B,j,n\ra}\right)\,.
\end{aligned}
\ee
This form of the integrand was found in~\cite{Arkani-Hamed:2010kv} using the loop-level BCFW recursion relation.

%%%%%%%%%%%%%%%%%%%%%%%%%%%%%%%%%%%%%%%%%%%%%%%%%%%
%%%%%%%%%%%%%%%%%%%%%%%%%%%%%%%%%%%%%%%%%%%%%%%%%%

\subsection{NMHV 1-loop}
\label{sec:NMHV1loop}

There are two classes of MHV diagram that contribute to the 1-loop planar NMHV amplitude, each shown in figure~\ref{fig:1loopNMHV}.

For the `triangle' topology, there must be at least one external leg at each vertex. Consequently, legs $i\!-\!1$, $j\!-\!1$ and $k\!-\!1$ are always attached to their respective propagators, so the momentum twistor expression for this diagram is
\be
	[\,\rf\,,i\!-\!1,i,A,B']\,[\,\rf\,,j\!-\!1,j,A,B'']\,[\,\rf\,,k\!-\!1,k,A,B''']
\label{NMHVtriangle}
\ee
where
\be
	B'=(A,B)\cap(\,\rf\,,k\!-\!1,k)\,,\quad B'' = (A,B)\cap(\,\rf\,,i\!-\!1,i)\,,\quad B''' = (A,B)\cap(\,\rf\,,j\!-\!1,j)
\ee
in accordance with the rule above.

\FIGURE[t]{
	\includegraphics[width=140mm]{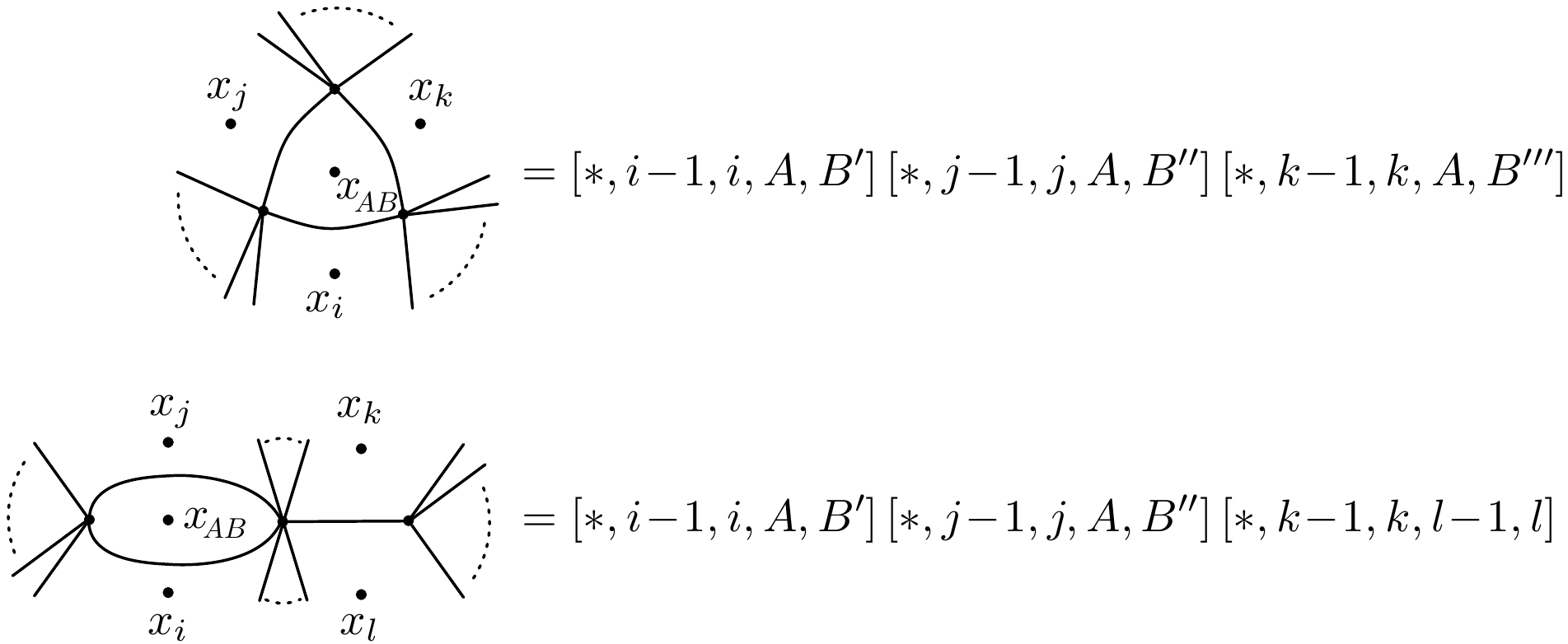}
	\caption{The two generic topologies of MHV diagram that contribute to the 1-loop planar NMHV amplitude.}
	\label{fig:1loopNMHV}
}

For the `bubble + leg' topology, in the generic case that $j\neq k$ and $l\neq i$ (mod $n$) the momentum twistor MHV rule assigns R invariants to the propagators independently, with the result that one obtains a factor of
\be
	[\,\rf\,,i\!-\!1,i,A,B']\,[\,\rf\,,j\!-\!1,A,B'']\times[\,\rf\,,k\!-\!1,k,l\!-\!1,l]
\ee
with $B'=(A,B)\cap(\,\rf\,,j\!-\!1,j)$ and $B''=(A,B)\cap(\,\rf\,,i\!-\!1,i)$. This is simply the product of a the expressions for a 1-loop MHV diagram and a tree-level NMHV diagram. However, there are also exceptional cases of this diagram where either $k=j$ (so that there are no external legs attached to the middle vertex `upstairs') or $l=i$, or both. The rule requires that we shift the external twistors in such cases, so that overall we find
\be
	[\,\rf\,,\widehat{i\!-\!1},i,A,B']\,[\,\rf\,,j\!-\!1,A,B'']\times[\,\rf\,,\widehat{k\!-\!1},k,l\!-\!1,l]
\label{NMHVbubbleleg}
\ee
where $B'=(A,B)\cap(\,\rf\,,j\!-\!1,j)$ and $B''=(A,B)\cap(\,\rf\,,\widehat{i\!-\!1},i)$ and where the shifted external twistors are
\be
\begin{aligned}
	\widehat{i\!-\!1} &= 
		\begin{cases}
			(i\!-\!1,i)\cap(\,*\,,k\!-\!1,k) & \hbox{if } l=i \hbox{ (mod } n)\\
			\ i\!-\!1 & \hbox{otherwise}
		\end{cases}
		\\
	\widehat{k-1} &=
		\begin{cases}
			(k\!-\!1,k)\cap(\,*\,,A,B'') & \hbox{if } k=j\\
			\ k\!-\!1 & \hbox{otherwise.}
		\end{cases}
\end{aligned}
\ee
Note that the shift of $i\!-\!1$ is itself shifted in the doubly exceptional case that there are no external legs attached to the middle vertex at all. The complete planar amplitude is then given by
\be
	M_{\rm NMHV}^{(1)} = \int {\rm D}^{3|4}Z_A\wedge{\rm D}^{3|4}Z_B\ 
	\left(\sum_{i<j<k} \hbox{(triangle)} + \sum_{i<j\leq k<l }\hbox{(bubble with leg)}\right),
\ee
where the limits on the sums are understood mod $n$. Rather than examining this amplitude in isolation, we proceed to the general proof at 1-loop.

%%%%%%%%%%%%%%%%%%%%%%%%%%%%%%%%%%%%%%%%%%%%%%%%%%%
%%%%%%%%%%%%%%%%%%%%%%%%%%%%%%%%%%%%%%%%%%%%%%%%%%

\subsection{General 1-loop integrands}
\label{sec:gen1loop}

We now give a general proof of that the one-loop momentum twistor MHV rules stated above agree with the standard MHV rules in momentum space. That these rules in turn agree with the Feynman diagram calculation was proved in~\cite{Brandhuber:2005kd}. 

\medskip

The treatment of the propagators that do not participate in the loop, and the shifts of external momentum twistors in the boundary cases proceeds exactly as at tree level. It remains to prove that our rule for the assignment of R invariants to loop propagators agrees with the usual momentum space rules. To handle these, number the vertices around the (planar) loop clockwise from $1$ to $m$, and suppose that external legs $j_r$ to $j_{r+1}-1$ are attached to the $r^{\rm th}$ such vertex. The loop variables may be handled by the usual CSW prescription, which associates the spinor $|\ell_r\ra\equiv  (x_0-x_{j_r})|\iota]$ to each propagator in the loop. Similarly, the sum over the possible members of the $\cN=4$ supermultiplet flowing along the propagator separating $x_0$ from$x_{j_r}$ may be associated with an integral over a fermionic variable $\eta_r$. The $m$ vertices in the loop then contribute a product 
\be
\begin{aligned}
	&\prod_{r=1}^m \delta^{0|8}(\theta_{j_{r+1},j_r} +|\ell_{r+1}\ra \eta_{r+1}-|\ell_r\ra \eta_r)\\
	&\ =\ \delta^{0|8}(\theta_{j_{m+1}}-\theta_{j_1})
	\prod_{r=1}^{m-1} \delta^{0|8} (\theta_{j_{r+1},j_r} +|\ell_{r+1}\ra\eta_{r+1}-|\ell_r\ra \eta_r)
\end{aligned}
\label{loopGrassmannDeltas}
\ee
of $m$ Grassmann $\delta^{0|8}$-functions. In the second line, we have substituted the arguments of the first $m-1$ such $\delta$-functions into the arguments of the $m^{\rm th}$ one, whereupon it becomes simply an overall fermionic momenta conserving $\delta$-function for the loop. This prefactor will be left as part of the overall MHV factor associated to the loop part of the diagram.

Just as the standard bosonic region momentum $x_0$ for the loop trivialises momentum conservation for the whole loop, so too can we trivialise fermionic momentum conservation within the loop by defining a region supermomentum $\theta_0$ by
\be
	\theta_0\equiv \theta_{j_m} + |\ell_m\ra \eta_m
\ee
and multiplying~\eqref{loopGrassmannDeltas} by the additional factor $\delta^{0|8}(\theta_{j_m,0} + |\ell_m\ra \eta_m)$, and extending the loop integral as $\rd^4 x_0\longrightarrow \rd^{4|8}x_0$. On the support of this extra $\delta^{0|8}$-function, the other Grassmann $\delta$-functions can be reduced inductively to enforce
\be
	\theta_0-\theta_{j_r} = |\ell_r\ra \eta_r\, ,
\ee
so that~\eqref{loopGrassmannDeltas} is equivalent to the product 
\be 
	\delta^{0|8}(\theta_{j_{m+1}}-\theta_{j_1})\,
	\prod_{r=1}^{m} \delta^{0|8} (\theta_{j_r,0} +|\ell_r\ra\eta_{r})\, .
\ee
We can now integrate out the fermionic variables $\eta_r$ for each propagator in the loop. The result is
\be
	\delta^{0|8}(\theta_{j_{m+1}}-\theta_{j_1})\,
	\prod_{r=1}^{m} \delta^{0|4} (\la\ell_r |\theta_{j_{r},0}\ra)\,,
\ee
enforcing overall supermomentum conservation for the loop together with a
factor of $\delta^{0|4}(\la\ell_r|\theta_{j_{r},0}\ra)$ for the $r^{\rm th}$ propagator.  We now pull out an overall Parke-Taylor denominator for the external lines attached to the loop, together with the overall super-momentum conserving factor $\delta^{0|8}(\theta_{j_{m+1}}-\theta_{j_1})$ for the loop. Including a $1/p^2$ factor for each propagator, we find the product
\be
\label{1loop-contrib}
	\prod_{r=1}^m \frac{\la j_r\, j_r\!-\!1\ra \ \delta^{0|4}\!\left(\la\ell_r|\theta_{j_{r}0}\ra\right)}
	{x_{j_r,0}^2\, \la\ell_r\,j_r\ra\,\la\ell_r\,j_r\!-\!1\ra\,\la\ell_r\,\ell_{r+1}\ra}\, .
\ee
from the usual MHV rules in momentum twistor space, where the loop integral is to be taken over the auxiliary region supermomentum $(x_0,\theta_0)$. 

\medskip

We must now translate this expression to momentum twistor space and compare it with the product of R-invariants provided by our momentum twistor rule.   To do so, we introduce the momentum twistors 
\be
\begin{aligned}
	A &\equiv \big(\la \lambda_A,\,\la \lb_A |x_0\,,\, \la \lb_A |\theta_0\big)\\
	B_r &\equiv \big(\la\ell_r|,\, \la\ell_r|x_0\,,\,  \la\ell_r|\theta_0\big) 
\end{aligned}
\ee
where $\la \lb_A|$ is an arbitrary spinor, subject only to the conditions $\la \lb_A|\ell_r\ra\neq 0$ for all $r$. Clearly, the twistor lines $(A,B_r)$ coincide with the line ${\rm X}_0$ representing the auxiliary region $(x_0,\theta_0)$.  Geometrically, as in figure~\ref{fig:transversal} at tree-level, the point $B_r$ represents the intersection of ${\rm X}_0$ with the unique transversal through ${\rm X}_0$, ${\rm X}_{j_r}$ and the reference twistor $Z_\rf$. It is therefore equal to the shifted momentum twistor $\widehat{B}$ that was earlier associated to the $(r+1)^{\rm st}$ loop propagator.

Upon inserting factors into~\eqref{1loop-contrib} that cancel around the loop, we obtain the
equivalent expression\footnote{As at tree-level, by a small modification of the choice of factor included in~\eqref{1loop-contrib2} we can produce an equivalent rule for the opposite orientation of the loop.}
\be
	\prod_{r=1}^m \frac{\la\ell_{r-1}\,\lb_A\ra}{\la\ell_r\,\lb_A\ra}
	\times
	\frac{\la j_r\, j_r\!-\!1\ra \ \delta^{0|4}\!\left(\la\ell_r|\theta_{j_{r}0}\ra\right)}
	{x_{j_r,0}^2\, \la\ell_r\,j_r\ra\,\la\ell_r\,j_r\!-\!1\ra\,\la\ell_r\,\ell_{r+1}\ra}\, ,
\label{1loop-contrib2}
\ee
(where of course $|\ell_0\ra\equiv|\ell_m\ra$). It is straightforward to recognise this as the product of R-invariants
\be
	\prod_{r=1}^m\,  [\,\rf\,,j_r\!-\!1,j_r, A,B_{r-1}]\,,
\ee
exactly in agreement with the momentum twistor MHV rule given at the beginning of the section, for the specific choice of reference twistor $Z_\rf = (0,\iota,0)$. As at tree level, the proof~\cite{Brandhuber:2005kd} that the momentum space rules are equivalent to the usual Feynman rules ensures that the complete integrand is independent of $|\iota]$. Since this integrand is dual superconformally invariant, we can treat $Z_\rf$ as a completely arbitrary reference momentum supertwistor. We have thus proved the validity of the momentum twistor MHV rules for 1-loop amplitudes.

%%%%%%%%%%%%%%%%%%%%%%%%%%%%%%%%%%%%%%%%%%%%%%%%%%%
%%%%%%%%%%%%%%%%%%%%%%%%%%%%%%%%%%%%%%%%%%%%%%%%%%

\subsection{The loop contour}
\label{sec:loopintegral}

At one loop, the momentum twistor MHV rules have provided us with expressions for the integrands of planar amplitudes in terms of a dual superconformally invariant sum of products of R invariants. In the above, we stated that to perform the loop integrals themselves, one must integrate this expression over the complete auxiliary supertwistors $Z_A$ and $Z_B$ using the measure ${\rm D}^{3|4}Z_A\wedge{\rm D}^{3|4}Z_B$. Let us  see why this is correct.

Whenever $A$ and $B$ are distinct points in projective space, they define a line X that corresponds to an arbitrary point $x_{AB}$ in (complexified) space-time. Explicitly, we have\footnote{The subscripts in this equation refer to points $A$ and $B$, and not spinor components!}
\be
	x_{AB} = \frac{\mu_B\lambda_A - \mu_A\lambda_B}{\la A\,B\ra}
	\qquad
	\theta_{AB} = \frac{\chi_B\lambda_A - \chi_A\lambda_B}{\la A\,B\ra}
\ee
which ensure that $A,B\in {\rm X}$. Using this, our integration measure may be split as
\be
	{\rm D}^{3|4}Z_A\wedge{\rm D}^{3|4}Z_B 
	= \frac{\la \lb_A\rd\lb_A\ra\wedge\la\lb_B\rd\lb_B\ra}{\la\lb_A\,\lb_B\ra^2}\wedge\rd^4x_{AB}\,\rd^8\theta_{AB}
\label{measuredecomposition}
\ee
into a measure on the choice of line $(x_{AB}, \theta_{AB})$ together with a measure for the actual locations of $A$ and $B$ on this line. These locations are not part of the original momentum space data, and so must drop out.

To understand how this happens, recall that by our above proof the sum of products of R invariants in the integrand is equivalent to the standard, Feynman diagram formula for the loop integrand. After integrating out the fermions, this sum can only be a function of the region momentum $x_{AB}$ and the external twistors. Therefore, the only dependence on $\lambda_{A,B}$ in the entire integral is from the measure~\eqref{measuredecomposition}. It is now straightforward to integrate out these spinors. Each $\lambda$ represents a point on the line ${\rm X}\cong\CP^1$. The integrand depends on $\lambda_{A,B}$ meromorphically, so the integral must be treated as a contour integral. The only class of  contour that leads to a non-vanishing result is to pick the contour to be the antidiagonal $\overline{\CP^1}\subset\CP^1\times\CP^1$. 
This antidiagonal is defined by setting $\lambda_B$ to be the complex conjugate of $\lambda_A$, using any notion of spinor complex conjugation that has no fixed points. A spinor complex conjugation with no fixed points (such as Euclidean conjugation) has the property that if $\lambda_A = \overline{\lambda}_B$, then  $\lambda_A\neq\lambda_B$ ensureing that the contour avoids the singularity. With any such contour, the $\lambda_{A,B}$ integrand becomes simply $2\pi\im$ times the K\"ahler form $\omega$ on $S^2$ so the integral may be trivially performed:
\be
	\oint_{\overline{\lb}_A=\lb_B} \frac{\la\lb_A\rd\lb_A\ra\wedge\la\lb_B \rd\lb_B\ra}{\la\lb_A\,\lb_B\ra^2} 
			= \int_{S^2} \frac{\la\lb_A\rd\lb_A\ra\wedge\overline{\la\lb_A \rd\lb_A\ra}}{\la\lb_A\,\overline{\lb}_A\ra^2}
			= 2\pi\im \int_{S^2}\omega = 2\pi\im
\ee
leaving only the usual loop integral $\rd^4x_{AB}$.

While this explains how to reduce the twistor integrals to the usual momentum space integral, note that it is also possible to perform the integrals in twistor space directly. Na{\" i}vely, in Euclidean signature, or any Wick rotation away from the Euclidean real slice towards the Lorentzian one, the total integral is over the $S^2$ contour of the $\lambda$ integrals and the $\mathbb{R}^4$ loop integral, viewed as a real contour inside complex space-time. Since the integrand is dual conformally invariant, it is natural to extend these integrations over the compact space. In so doing, one finds that rather than obtaining a product $S^2\times S^4$, the $S^2$ factor fibres over the $S^4$ base so that the total contour is the antidiagonal $\overline{\CP}^3\subset\CP^3\times\CP^3$ inside the two copies of twistor space parametrised by (the bosonic components of) $Z_A$ and $Z_B$ (see~\cite{Hodges:2010kq,Mason:2010pg} for further discussion). This na{\" i}ve extension fails because of  the IR divergences of the loop amplitudes -- there is no way to make such a contour avoid the singularities of the integrand. However, it is possible to regularise the integrand directly in twistor space using the Coulomb branch regularisation introduced in~\cite{Alday:2009zm} and this has been carried out explicitly in~\cite{Hodges:2010kq,Mason:2010pg} for 1-loop and~\cite{Drummond:2010mb} for certain 2-loop integrals. (At present, this regularisation technique requires that the integrand be written in a way that contains no spurious loop propagators.)

%%%%%%%%%%%%%%%%%%%%%%%%%%%%%%%%%%%%%%%%%%%%%%%%%%%
%%%%%%%%%%%%%%%%%%%%%%%%%%%%%%%%%%%%%%%%%%%%%%%%%%

\section{Relation to Generalised Unitarity}
\label{sec:boxes}

This section is an interlude from the main development of the paper. In it, we relate the form
\be
	M_{\rm MHV}^{(1)} = \int{\rm D}^{3|4}Z_A\wedge{\rm D}^{3|4}Z_B 
	\sum \,[\,\rf\,,i\!-\!1,i,A,B']\,[\,\rf\,,j\!-\!1,j,A,B'']
\label{MHVrules}
\ee
of the integrand of the 1-loop MHV amplitude provided by the MHV rules in~\eqref{MHV1loop} to the generalised unitarity form~\cite{Bern:1994zx}
\be
	M_{\rm MHV}^{(1)} = \sum F_{n;ij}\,,
\label{sumboxes}
\ee
where $F_{n;ij}$ is understood to mean either the `two-mass easy' box function 
\be
\begin{aligned}
	F^{2{\rm me}}_{n;ij}&\equiv \int \rd^4x_0 \ 
	\frac{x^2_{i,j}x^2_{i+1,j+1} - x^2_{i,j+1}x^2_{i+1,j}}{x_{0,i}^2 x_{0,i+1}^2 x_{0,j}^2 x_{0,j+1}^2}\\
	&= \int {\rm D}^3Z_A\wedge {\rm D}^3Z_B\  
	\frac{ \la j\!-\!1,j,j\!+\!1,i\ra\, \la j,i\!-\!1,i,i\!+\!1\ra}
	{\la A,B,i\!-\!1,i\ra\,\la A,B,i,i\!+\!1\ra\,\la A,B,j\!-\!1,j\ra\,\la A,B,j,j\!+\!1\ra}\\
\end{aligned}
\label{2me}
\ee
or the `one-mass' box function 
\be
\begin{aligned}
	F^{1{\rm m}}_{n;i} &\equiv \int \rd^4x_0 \ 
	\frac{x^2_{i-2,i}\,x^2_{i-1,i+1}}{x_{0,i-2}^2 x_{0,i-1}^2 x_{0,i}^2 x_{0,i+1}^2}\\
	&= \int {\rm D}^3Z_A\wedge {\rm D}^3Z_B\  
	\frac{\la i\!-\!3,i\!-\!2,i\!-\!1,i \ra\,\la i\!-\!2,i\!-\!1,i,i\!+\!1\ra}
	{\la A,B,i\!-\!3\,i\!-\!2\ra\,\la A,B,i\!-\!2,i\!-\!1\ra\,\la A,B,i\!-\!1,i\ra\,\la A,B,i,i\!+\!1\ra}\, ,
\end{aligned}
\label{1m}
\ee
which is just a special case of~\eqref{2me} with $j=i\!-\!2$. These integrals are divergent (on the loop contour) and must be regularised. As mentioned above, in the context of momentum twistors, it is natural to use the Coulomb branch regularisation of~\cite{Alday:2009zm} which amounts to replacing the propagators in~\eqref{2me}-\eqref{1m} according to the prescription
\be
 	\frac{1}{x_{0,i}^2} \longrightarrow \frac{1}{x_{0,i}^2 + \mu^2}
\ee
with some mass scale $\mu$ (related to the expectation value of the Coulomb branch scalar). This regularisation has a natural geometric interpretation in momentum twistor space, for which we refer the reader to~\cite{Hodges:2010kq,Mason:2010pg}.

Using dimensional regularisation and performing the loop integrals, the equality of the 1-loop MHV amplitude as computed by momentum space MHV rules with the generalised unitarity expression was famously proved in~\cite{Brandhuber:2004yw}, using an identity involving nine different dilogarithms. However, one message of the beautiful recent paper~\cite{Goncharov:2010jf} is that polylogarithm identities are best seen at the level of the integrand. Therefore, in this section we compare the two expressions~\eqref{MHVrules} and~\eqref{sumboxes} directly at the level of their integrands.

\begin{figure}[t]
	\centering
	\includegraphics[width=135mm]{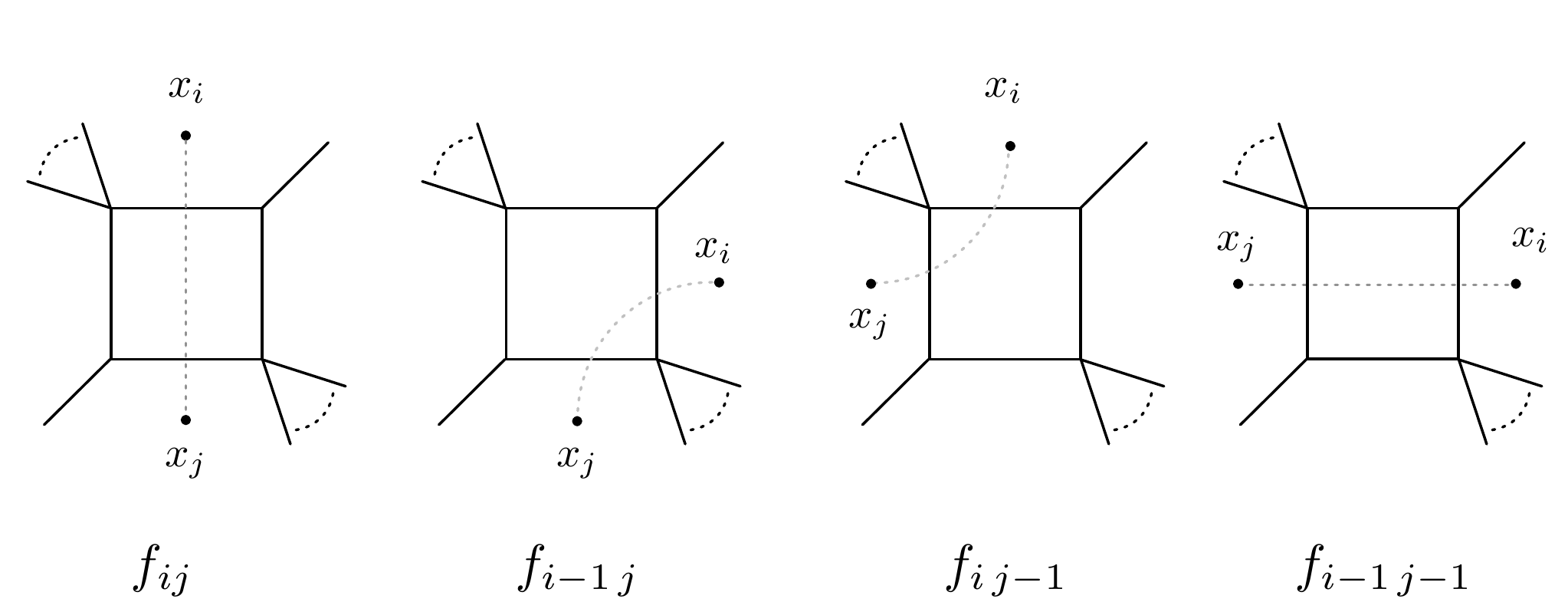}
	\caption{The four 2-cuts of two-mass easy boxes corresponding to the four terms in the expansion the 1-loop MHV 	diagram containing physical propagators $1/(x-x_i)^2$ and $1/(x-x_j)^2$.}
\label{fig:boxcuts}
\end{figure}

Consider the MHV diagram shown in figure~\ref{fig:1loopMHV} where external states $\{i,\ldots,j\!-\!1\}$ are attached to the left vertex and $\{j,\ldots,i\!-\!1\}$ to the right, and let us temporarily assume that there are at least two external particles on each vertex. As found in~\eqref{MHV1loopmomtw}, after integrating out the loop fermions, the integrand associated to this diagram is
\begin{multline}	
	\frac{\left(\la\,*\,,i\!-\!1,i,\big[A\ra,\la B\big],j\!-\!1,j,\,\rf\,\ra\right)^2}
	{\la A,B,i\!-\!1,i \ra\,\la A,B,j\!-\!1,j \ra\,\la A,B,i\!-\!1,\,\rf\,\ra\,\la A,B,i,\,\rf\,\ra\,\la A,B, j\!-\!1,\,\rf\,\ra\,\la A,B,j,\,\rf\,\ra}\\
	=\frac{\la\,\rf\,,j\!-\!1,j,\big[i\!-\!1\ra,\la i \big],A,B,\,\rf\,\ra\,\la\,\rf\,,i\!-\!1,i,\big[j\!-\!1\ra\,\la j\big],A,B,\,\rf\,\ra}
	{\la A,B,i\!-\!1,i \ra\,\la A,B,j\!-\!1,j \ra\,\la A,B,i\!-\!1,\,\rf\,\ra\,\la A,B,i,\,\rf\,\ra\,\la A,B, j\!-\!1,\,\rf\,\ra\,\la A,B,j,\,\rf\,\ra}\,,
\label{MHVrewrite}
\end{multline}		
where in going the second line we (twice) used the fact that
\be
	\la a,b,c,d\ra Z_e + \hbox{cyclic} = 0 
\ee
(four generic points form a basis of $\CP^3$) to rewrite the numerator. The second line of~\eqref{MHVrewrite} may be expanded out to give the sum of four terms
\be
	f_{i,j} + f_{i-1,j} + f_{i,j-1} + f_{i-1,j-1}
\ee
defined by
\be
\begin{aligned}
	f_{i,j} &= \frac{\la \,\rf\,,i,j\!-\!1,j\ra\la \,\rf\,,j,i\!-\!1,i \ra}{\la A,B,i,\,\rf\,\ra\,\la A,B,j,\,\rf\,\ra\,\la A,B,i\!-\!1,i\ra\,\la A,B,j\!-\!1,j\ra}\\
	f_{i-1,j} &= \frac{\la\,\rf\,,i\!-\!1,j\!-\!1,j\ra\,\la j,i\!-\!1,i,\,\rf\,\ra}{\la A,B,i\!-\!1,\,\rf\,\ra\,\la A,B,j,\,\rf\,\ra\,\la A,B,i\!-\!1,i\ra\,
			\la A,B,j\!-\!1,j\ra}\\
	f_{i,j-1} &= \frac{\la \,\rf\,,i,j\!-\!1,j\ra\,\la j\!-\!1,i\!-\!1,i,\,\rf\, \ra}{\la A,B,i,\,\rf\,\ra\,\la A,B,j\!-\!1,\,\rf\,\ra\la A,B,i\!-\!1,i\ra\,
			\la A,B,j\!-\!1,j\ra}\\
	f_{i-1,j-1} &= \frac{\la \,\rf\,,i\!-\!1,j\!-\!1,j\ra\,\la \,\rf\,,j\!-\!1,i\!-\!1,i \ra}{\la A,B,i\!-\!1,\,\rf\,\ra\,\la A,B,j\!-\!1,\,\rf\,\ra\,
			\la A,B,i\!-\!1,i\ra\,\la A,B,j\!-\!1,j\ra}
\end{aligned}
\label{fijdefined}
\ee
Each of these functions is associated to a 2-cut of a 2me box in the channel $x_{ij}^2\to0$, but different $f$s correspond to cuts of different boxes (see figure~\ref{fig:boxcuts}). Thus, a single MHV diagram generically contributes to four different 2me boxes (although some of these vanish when $|i-j|< 2$ mod $n$).  

\begin{figure}[t]
	\flushleft
	\includegraphics[width=152mm]{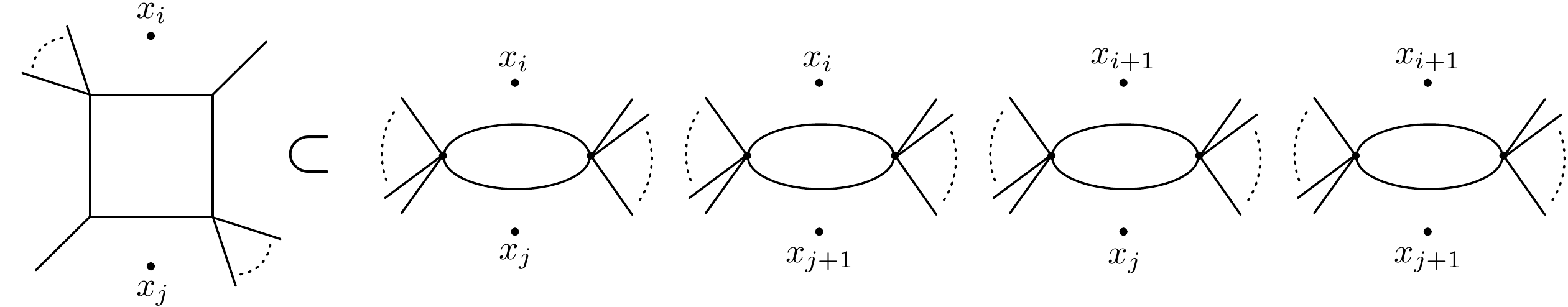}
	\caption{The two mass easy box function $F_{n;ij}^{\rm 2me}$ receives contributions from the four
	displayed MHV diagrams.}
\label{fig:MHVcuts}
\end{figure}

Conversely, a generic 2me box $F^{2{\rm me}}_{n;ij}$ receives contributions from four different 1-loop MHV diagrams, shown in figure~\ref{fig:MHVcuts}. Adding together the $f$s that correspond to 2-cuts of the \emph{same} 2me box integral gives 
\be
\begin{aligned}
	&\frac{\Big\la (\rf \,,i,j)\!\cap\!(j\!-\!1, j\!+\!1),j,A,B\Big\ra \times \Big\la (\rf\,,i,j)\!\cap i\!-\!1, i\!+\!1),i,A,B\Big\ra }
	{\la A,B,i,\,\rf\,\ra\,\la A,B,j,\,\rf\,\ra\la A,B,i\!-\!1,i\ra\,\la A,B,i,i\!+\!1\ra\,\la A,B,j\!-\!1,j\ra\,\la A,B,j,j\!+\!1\ra}\\
	&= \ \frac{\Big\la (\rf\,,i)\!\cap\!(j\!-\!1,j,j\!+\!1),j,A,B\Big\ra \times \Big\la (\rf\,,j)\!\cap\!(i\!-\!1,i,i\!+\!1),i,A,B\Big\ra}
	{\la A,B,i,\,\rf\,\ra\,\la A,B,j,\,\rf\,\ra\la A,B,i\!-\!1,i\ra\,\la A,B,i,i\!+\!1\ra\,\la A,B,j\!-\!1,j\ra\,\la A,B,j,j\!+\!1\ra}
\end{aligned}
\label{sum4MHVs}
\ee
where the first line comes directly from the sum of $f$s. The second line follows from the first because point $(\rf\,,i)\!\cap\!(j\!-\!1,j,j\!+\!1)$ differs from point $(\rf \,,i,j)\!\cap\!(j\!-\!1, j\!+\!1)$ only by a translation in the direction of $j$, which vanishes in the skew product. Similarly, point $(\rf\,,j)\!\cap\!(i\!-\!1,i,i\!+\!1)$ differs from $(\rf\,,i,j)\!\cap( i\!-\!1, i\!+\!1)$ only by a translation in the direction of $i$. If we call these twistors
\be
	I_* \equiv (\rf\,,i)\!\cap\!(j\!-\!1,j,j\!+\!1)
	\qquad\hbox{and}\qquad
	J_* \equiv (\rf\,,j)\!\cap\!(i\!-\!1,i,i\!+\!1)\,,
\ee
then the numerator of~\eqref{sum4MHVs} may be written as
\be
\begin{aligned}
	&\la A,B, I_*,j\ra\,\la A,B,J_*,i\ra \\
	&\ = \la A,B,I_*,i\ra\,\la A,B,J_*,j\ra + \la A,B,I_*,J_*\ra\,\la A,B,i,j\ra\\
	&\ =\la A,B,i,*\ra \la A,B,j,*\ra \la i,j\!-\!1,j,j\!+\!1\ra \la j,i\!-\!1,i,i\!+\!1\ra + \la A,B,I_*,J_*\ra\,\la A,B,i,j\ra\ .
\end{aligned}
\ee
In the final line, the first term cancels the spurious propagators $\la A,B,i,*\ra\,\la A,B,j,*\ra$ in the denominator of~\eqref{sum4MHVs}, leaving us with precisely the 2 mass easy integrand
\be
	\frac{ \la j\!-\!1,j,j\!+\!1,i\ra\, \la j,i\!-\!1,i,i\!+\!1\ra}
	{\la A,B,i\!-\!1,i\ra\,\la A,B,i,i\!+\!1\ra\,\la A,B,j\!-\!1,j\ra\,\la A,B,j,j\!+\!1\ra}\,.
\ee
The second term combines with the denominator to form
\be
	\frac{\la A,B,I_*,J_*\ra\,\la A,B,i,j\ra}{\la A,B,i,\,\rf\,\ra\,\la A,B,j,\,\rf\,\ra}
	\times\frac{1}{\la A,B,i\!-\!1,i\ra\,\la A,B,i,i\!+\!1\ra\,\la A,B,j\!-\!1,j\ra\,\la A,B,j,j\!+\!1\ra}\, .
\ee
This term contains a product of four physical propagators, together with a prefactor of vanishing homogeneity in the loop twistors $A,B$. Aside from the analytic derivation above, we have checked numerically that this second term is independent of the reference twistor $Z_\rf$ for up to twelve external particles, when summed over  $1\leq i<j\leq n$. This prefactor is constructed from the lines $(A,B)$, $(i,*\,)$ and $(j,*\,)$, but also from the \emph{planes} $(i\!-\!1,i,i\!+\!1)$ and $(j\!-\!1,j,j\!+\!1)$, entering in the definition of $I_*$ and $J_*$. Dependence on planes is a signal that this term is parity odd. It therefore integrates to zero on any contour (such as the loop contour) that respects parity, although it can contribute to leading singularities.

We have thus shown that the sum of MHV diagrams reproduces the generalised unitarity sum of 2 mass easy boxes \emph{at the level of the integrand}, up to parity odd terms that integrate to zero on the loop contour.

%%%%%%%%%%%%%%%%%%%%%%%%%%%%%%%%%%%%%%%%%%%%%%%%%%%
%%%%%%%%%%%%%%%%%%%%%%%%%%%%%%%%%%%%%%%%%%%%%%%%%%

\section{Momentum twistor MHV rules to all loops}
\label{sec:allloops}

In this section we extend the momentum twistor MHV rules stated above to all planar diagrams in $\cN=4$ SYM.  The extension is a straightforward generalisation of the 1-loop case:
\begin{itemize}
	
	\item To every loop in the planar diagram, associate two auxiliary momentum twistors $A_m$ and $B_m$ 
		(with that $m=1,\ldots,L$).

	\item To every propagator in the planar diagram, associate an R invariant $[\,\rf\,,\ ,\ ,\ ,\ ]$ depending on an
		arbitrary reference momentum twistor $Z_\rf$ and four other momentum twistors.
			
	\item If a propagator separates the external region $x_i$ from the external region $x_j$, the remaining arguments are
		$\widehat{i\!-\!1}$, $i$, $\widehat{j\!-\!1}$ and $j$. The twistor $\widehat{Z_{i-1}}=Z_{i-1}$ if the propagator
		shares a vertex with external leg $i\!-\!1$. Otherwise, $\widehat{i\!-\!1}$ is the intersection of the line $(i\!-\!1,i)$		with the plane through the reference twistor $Z_\rf$ and the line associated to the preceding propagator in the
		cyclic ordering. This line may correspond to either an external region or a loop region. ($\widehat{j\!-\!1}$ 
		is determined similarly.)
		
	\item If a propagator separates the external region $x_i$ from the loop region $x_{(AB)_m}$, the remaining 		arguments are $\widehat{i\!-\!1}$, $i$, $A_m$ and $\widehat{B_m}$, where $\widehat{B_m}$ is the intersection
		of the line $(AB)_m$ with the plane spanned by $Z_\rf$ and the line $(AB)_{m-1}$.
	
	\item If a propagator separates the loop region $x_{(AB)_m}$ from the loop region $x_{(AB)_n}$, the remaining 
		arguments are $A_m$, $\widehat{B_m}$, $A_n$ and $\widehat{B_n}$. Again, $\widehat{B_m}$ is the
		intersection of the line $(AB)_m$ with the plane spanned by $Z_\rf$ and the line $(AB)_{m-1}$ (and
		similarly for $\widehat{B_n}$).
	
	\item Sum the expressions obtained as above from all contributing MHV diagrams, and integrate the result 
		using the holomorphic measure $\prod_{m=1}^L {\rm D}^{3|4}Z_{A_m}\wedge{\rm D}^{3|4}Z_{B_m}$ and the 
		$L$-fold product of the antidiagonal $\CP^3$ contour described in section~\eqref{sec:loopintegral} 
		(after regularising).

\end{itemize}

These rules may be proved by induction on loop order.  We start by considering a 1-loop subdiagram with $P_m$ propagators around the loop. By the results of section~\ref{sec:gen1loop} we may replace this by a product of $P_m$ R-invariants, times an overall factor of the MHV tree  -- including fermionic momentum conservation -- for the given subloop. The only difference compared to the 1-loop case is that the arguments of these R invariants may now be associated with further loop regions. (Since these regions are external to the loop under consideration, the 1-loop proof goes through unaltered for this sub-loop.) Setting the R invariants to one side, we may replace our original MHV diagram by one in which the whole sub-loop is replaced by a single MHV vertex. This new MHV diagram clearly has one fewer loops, allowing us to recurse down to the one-loop case.

%%%%%%%%%%%%%%%%%%%%%%%%%%%%%%%%%%%%%%%%%%%%%%%%%%%
%%%%%%%%%%%%%%%%%%%%%%%%%%%%%%%%%%%%%%%%%%%%%%%%%%

\subsection{The 2-loop MHV amplitude}
\label{sec:2loopMHV}

\begin{figure}
	\centering
	\includegraphics[height=38mm]{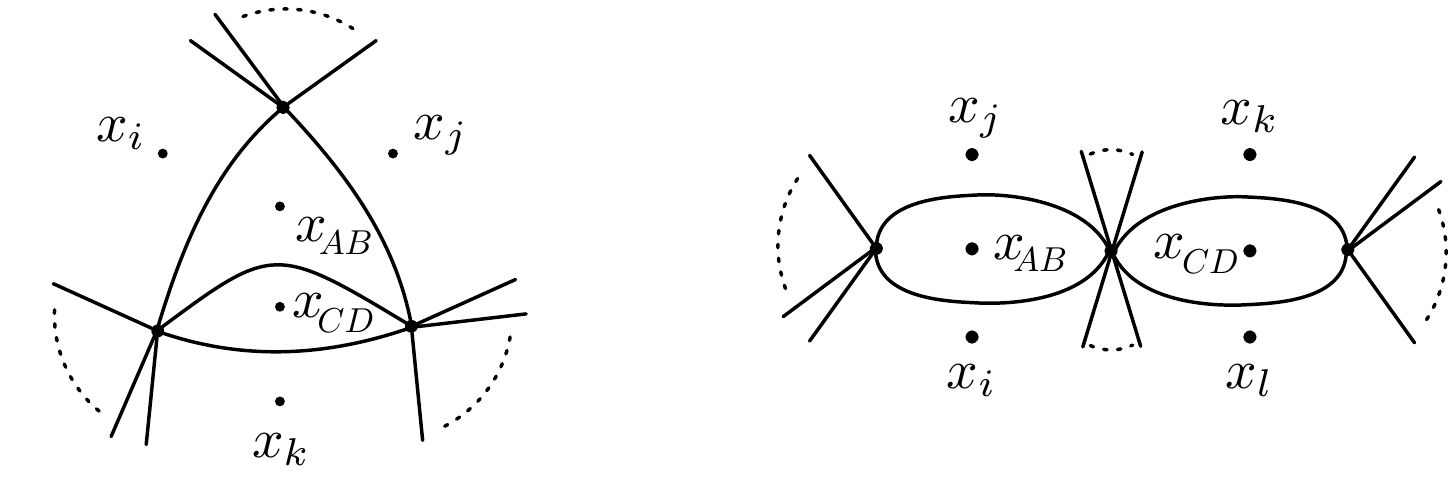}
	\caption{The two generic topologies of MHV diagram that contribute to the planar 2-loop MHV amplitude.}
	\label{fig:2loop}
\end{figure}

We finish by illustrating the above all-loop momentum twistor MHV rule by computing the integrand of the $n$-particle 2-loop planar MHV amplitude. There are two generic classes of MHV diagram, shown in figure~\ref{fig:2loop}. Applying the MHV rules to these diagrams gives
\be
	M_{\rm MHV}^{(2)} = M_1 + M_2
\ee
where 
\be
	M_1 = \int_{ABCD} \sum\,[\,\rf\,,\widehat{i\!-\!1},i,A,B']\,[\,\rf\,,j\!-\!1,j,A,B'']\,[\,\rf\,,A,B''',C,D'']\,[\,\rf\,,\widehat{k\!-\!1},k,C,D']
\ee
corresponds to the sum of `triangle-bubbles'. Here,  the loop shifts are defined by
\be
\begin{aligned}
	&B' &= &\   (A,B)\cap(\,\rf\,,C,D) \qquad\qquad 	&D' &=\ (C,D)\cap(\,\rf\,,A,B)\\
	&B'' &= &\ (A,B)\cap(\,\rf\,,i\!-\!1,i) 			&D'' &= \ (C,D)\cap(\,\rf\,,k\!-\!1,k)\\
	&B''' &= &\  (A,B)\cap(\,\rf\,,j\!-\!1,j) & &
\end{aligned}
\ee
while the external shifts are
\be
\begin{aligned}
	\widehat{i\!-\!1} &= 
		\begin{cases}
			(i\!-\!1,i)\cap(\,\rf\,,C,D) 	& \hbox{if } k=i\\
			\ i\!-\!1 				& \hbox{otherwise}
		\end{cases}
	\\
	\widehat{k\!-\!1} &=
		\begin{cases}
			(k\!-\!1,k)\cap(\,\rf\,,A,B)	& \hbox{if } j=k\\
			\ k\!-\!1				& \hbox{otherwise.}
		\end{cases}
\end{aligned}
\ee
The summation range of this term is $1\leq i < j\leq k \leq i$, understood mod $n$. The second term, corresponding to the sum of `double-bubbles', is
\be
	M_1 = \int_{ABCD}\sum \,[\,\rf\,,\widehat{i-1},i,A,B']\,[*,j-1,j,A,B'']\,[\,\rf\,,\widehat{k-1},k,C,D']\,[*,l-1,l,C,D'']
\ee
and has the loop shifts
\be
\begin{aligned}
	&B' &= &\ (A,B)\cap(\,\rf\,,j\!-\!1,j) \qquad\qquad 	&D' &=\ (C,D)\cap(\,\rf\,,l\!-\!1,l)\\
	&B'' &= &\ (A,B)\cap(\,\rf\,,i\!-\!1,i) 			&D'' &=\ (C,D)\cap(\,\rf\,,k\!-\!1,k)
\end{aligned}
\ee
and the external shifts
\be
\begin{aligned}
	\widehat{i\!-\!1} &=
		\begin{cases}
			(i\!-\!1,i)\cap(\,\rf\,,C,D) 	& \hbox{if } l=i\\
			\ i\!-\!1 				& \hbox{otherwise}\\
		\end{cases}
	\\
	\widehat{k\!-\!1} &=
		\begin{cases}
			(k\!-\!1,k)\cap(\,\rf\,,A,B) 	& \hbox{if } k=j\\
			\ k\!-\!1			 	& \hbox{otherwise.}
		\end{cases}
\end{aligned}
\ee
The summation range in this second term is $1\leq i<j\leq k< l\leq i$, again understood mod $n$. In both $M_1$ and $M_2$ we have used to abbreviation $\int_{ABCD}$ to indicate the integral over the momentum twistors associated to each loop.

%%%%%%%%%%%%%%%%%%%%%%%%%%%%%%%%%%%%%%%%%%%%%%%%%%%
%%%%%%%%%%%%%%%%%%%%%%%%%%%%%%%%%%%%%%%%%%%%%%%%%%

\acknowledgments

It is a pleasure to thank Tim Adamo, Nima Arkani-Hamed, Rutger Boels, Jacob Bourjaily, Freddy Cachazo, Simon Caron-Huot and Jaroslav Trnka for many helpful discussions. The work of DS is supported by the Perimeter Institute for Theoretical Physics. Research at the Perimeter Institute is supported by the Government of Canada through Industry Canada and by the Province of Ontario through the Ministry of Research $\&$ Innovation. The work MB is supported by an STFC Postgraduate Studentship. The work of LM  was financed in part by EPSRC grant number EP/F016654.

\bibliographystyle{JHEP}
\bibliography{MHV}

\end{document}